\documentclass[twocolumn,amsmath,amssymb,prl,10pt,nofootinbib,superscriptaddress]{revtex4}
\usepackage{ graphicx, float,amsmath, amsmath, amssymb}
\usepackage[export]{adjustbox}

\def\be{\begin{equation}}
\def\ee{\end{equation}}
\def\bea{\begin{eqnarray}}
\def\eea{\end{eqnarray}}
\def\bse{\begin{subequations}}
\def\ese{\end{subequations}}

\newcommand{\sums}{\sum\limits}
\usepackage[breaklinks, colorlinks, citecolor=blue]{hyperref}

\usepackage[labelsep=period]{caption}
\usepackage[normalem]{ulem}
\usepackage{soul}
\usepackage{minitoc}
\usepackage{physics}
\usepackage{caption}

\usepackage{bm}

\begin{document}

\title{Detailed analysis of the curvature bounce:\\ background dynamics and imprints in the CMB}

\author{Cyril Renevey}%
\affiliation{%
Laboratoire de Physique Subatomique et de Cosmologie, Université Grenoble-Alpes, CNRS/IN2P3\\
53, avenue des Martyrs, 38026 Grenoble cedex, France
}

\author{Aurélien Barrau}
\affiliation{%
Laboratoire de Physique Subatomique et de Cosmologie, Université Grenoble-Alpes, CNRS/IN2P3\\
53, avenue des Martyrs, 38026 Grenoble cedex, France
}

\author{Killian Martineau}
\affiliation{%
Laboratoire de Physique Subatomique et de Cosmologie, Université Grenoble-Alpes, CNRS/IN2P3\\
53, avenue des Martyrs, 38026 Grenoble cedex, France
}

\date{\today}

\begin{abstract} 
If the spatial sections of the Universe are positively curved, extrapolating the inflationary stage backward in time inevitably leads to a classical bounce. This simple scenario, non-singular and free of exotic physics, deserves to be investigated in details. The background dynamics exhibits interesting features and is shown to be mostly insensitive to initial conditions as long as observational consequences are considered. The primordial scalar power spectrum is explicitly computed, for different inflaton potentials, and the subsequent CMB temperature anisotropies are calculated. The results are compatible with current measurements. Some deviations with respect to the standard paradigm can however appear at large scales and we carefully disentangle what is associated with the vacuum choice with what is more fundamentally due to the bounce itself.

\end{abstract}

\maketitle

\section{Introduction}

The question of the shape of the spatial sections of the Universe is an old one. It is still unanswered but it has attracted a lot of attention recently, based on the refined observations of the cosmic microwave background (CMB). The latest data released by the Planck collaboration \cite{Aghanim:2018eyx} might favor a positively curved universe, described by a curvature density $\Omega_K=-0.044$, with a high  confidence level \cite{Aghanim:2018eyx,Park:2017xbl,Handley:2019tkm,DiValentino:2019qzk}. Although quite convincing when using CMB data alone, this result is in tension with other measurements, such as the baryonic acoustic oscillations. It is also dependant upon specific statistical priors that can be questioned \cite{Efstathiou:2020wem}. Nonetheless, even ignoring the claims of \cite{DiValentino:2019qzk}, the possibility that the universe is positively curved is worth being phenomenologically considered as a possible situation. Especially when taking into account, as we shall see, that this might cure naturally the initial singularity problem. Not to mention that some -- speculative but reasonable -- arguments from quantum gravity also favor this possibility (a finite space is a natural infrared regulator). 

It is believed in the standard cosmological paradigm that at its earliest stage the universe grew exponentially fast in a quasi-de Sitter phase. While not fully confirmed, the theory of inflation is supported by strong observational evidences \cite{Planck:2018jri}, the most obvious being the slightly red-tilted spectrum of primordial fluctuations. Strong constraints on non-Gaussianities in the CMB \cite{Planck:2019kim} favor a single field scenario. 

In the inflationary model, the temperature anisotropies observed in the CMB are explained by quantum fluctuations of the inflaton scalar field. These initial fluctuations can be evolved using the theory of gauge-invariant linear perturbations \cite{Mukhanov:1990me} and the primordial power spectrum can then be unambiguously calculated. 
The study of scalar and tensor perturbations was carried out in the case of a closed universe in \cite{Bonga:2016iuf,Bonga:2016cje}. It was concluded that a smaller-than-usual power is to be expected at very large scales. 

Although straightforward to show, it is often forgotten that a de Sitter space-time with closed spatial sections evades the big bang singularity and instead, naturally leads to a bounce. The possible implementations of this scenario were studied in \cite{Martin:2003bp,Uzan:2003nk,Barrau:2020nek}. The singular origin of our universe is naturally regularized thanks to the positive curvature combined with inflation. A further numerical investigation was carried out in \cite{Renevey:2020zdj}. Quite impressively, the usual claims about the need of a quantum (or modified) theory of gravity to escape the unavoidable singularity are simply contradicted using only the hypotheses of the standard cosmological scenario (if space is positively curved). This happens without any exotic physics. In \cite{Renevey:2020zdj}, we have studied the primordial tensor power spectrum with quantum fluctuations originating prior to the bounce. No noticeable difference was found when compared to the no-bounce version of \cite{Bonga:2016cje}. However, the equation of motion for the scalar perturbations being very different from the one of the tensor modes, we must investigate the scalar sector to predict reliably the possible imprint of the model on the CMB temperature anisotropies. This is the main goal of this work which builds on our previous papers \cite{Barrau:2020nek,Renevey:2020zdj}. The key-point consists in disentangling the effects due to the curvature (as in \cite{Bonga:2016iuf}) from those due to the bounce itself.

This paper is organized as follows. First, we describe the behaviour of the background dynamics during the period of inflation, around the bounce, and during the pre-bounce contracting universe. We analyze in details the dependence of the background dynamics upon initial conditions. Two different inflationary models are considered : a massive scalar field -- which is disfavoured by data but useful for comparisons -- and the Starobinsky potential \cite{Starobinsky:1980te,Martin:2013tda,Chowdhury:2019otk}. Then, we introduce the theory of linear perturbations in curved space and discuss the meaning of the Bunch-Davies vacuum before and after the bounce. Finally, we move on to the calculation of the primordial power spectra of  scalar perturbations, considering different cases for the initial vacuum. We compare our results with previous works, mainly with \cite{Bonga:2016iuf}, and with the available CMB data. 

Planck units are used throughout the paper except otherwise stated.

\section{Background dynamics}

To describe the background behaviour, homogeneity and isotropy, together with the closure of the spatial sections, are assumed. The topology of space-time is therefore $\mathbb{R}\times\mathbb{S}^3$, where $\mathbb{S}^3$ represents a hypersphere. The spatial curvature parameter $K>0$ is related to the physical radius $r(t)$ of the 3-sphere by $r^2(t)=a^2(t)/K$, where $a(t)$ is the dimensionless scale factor. Under these hypotheses, the FLRW metric can be written as
\begin{align}
    \dd s^2=-\dd t^2+\frac{a^2(t)}{K}\left(\dd\chi^2+\sin^2(\chi) \dd\Omega^2\right).
    \label{eq:metric}
\end{align}
The matter content of the universe is represented by a perfect fluid of density $\rho$ and pressure $p$, such that the Einstein field equations lead to the Friedmann and Raychaudhury equations:
\begin{align}
    H^2&=\frac{8\pi}{3}\rho-\frac{K}{a^2},\label{eq:friedmann}\\
    \dot{H}&=-4\pi(\rho+p)+\frac{K}{a^2},\label{eq:raychaudhury}
\end{align}
where $H=\dot{a}/a$ is the Hubble parameter and the dot notation stands for the derivative with respect to the cosmic time. From Eq.~\eqref{eq:friedmann}, it can immediately be seen that during an inflationary period, described by a quasi-constant energy density $\rho$, the curvature term on the right hand side (RHS) increases when the scale factor decreases. This behaviour inevitably leads to a bounce, at the time when $H=0$, instead of a singularity. A detailed analysis of the basic ingredients of the curvature bounce is given in \cite{Barrau:2020nek,Renevey:2020zdj}. Once again, it is worth emphasizing that the curvature of the universe is not yet firmly measured and could actually be positive, negative or null. A recent study \cite{DiValentino:2019qzk}  claimed that the CMB measurements do favor a closed universe with $\Omega_K:=-K/(3\bar{H}^2)=-0.044$, $\bar{H}=54.4$ km/s/Mpc being the current Hubble parameter. While the validity of this result is strongly debated \cite{Efstathiou:2020wem}, we assume here that it is correct. Even if it is not, a positive curvature remains anyway a possible situation -- somehow favored by theoretical arguments and not excluded by data -- deserving attention.

During the inflationary period, the matter content is described by a scalar field $\phi$. In this work, we consider two commonly used potentials for the inflaton: the quadratic and Starobinsky potentials, respectively given by
\begin{align}
    V(\phi)=\frac{1}{2}m^2\phi^2\quad ;\quad V(\phi)=\frac{3M^2}{32\pi}\left(1-e^{-\sqrt{16\pi/3}\phi}\right)^2,
\end{align}
where $m$ and $M$ are free parameters that can be constrained by CMB measurements, as we shall see below. The Klein-Gordon equation in the homogeneous FLRW metric given Eq.(\ref{eq:metric}) reads
\begin{align}
    \ddot{\phi}+3H\dot{\phi}+\pdv{V(\phi)}{\phi}=0.\label{eq:klein-gordon}
\end{align}
The density and pressure of the inflaton field can be written as
\begin{align}
    \rho=\frac{1}{2}\dot{\phi}^2+V(\phi)\quad \textrm{and}\quad p=\frac{1}{2}\dot{\phi}^2-V(\phi).\label{eq:rho_p}
\end{align}
As well known, Eqs. \eqref{eq:friedmann}, \eqref{eq:raychaudhury} and \eqref{eq:klein-gordon} are not independent. 

To perform the simulation, the initial conditions (IC) for the background are set at the bounce, which is chosen to be the origin of the time coordinate, $t=0$. This is only a matter of convenience and this does not mean that the system ``looses" memory of the pre-bounce phase. At the bounce, the derivative of the scale factor vanishes, $\dot{a}(0)=0$, and we choose, without any loss of generality, to set $a(0)=1$, fully fixing the IC for the homogeneous gravitational sector. In order to set the IC for the matter sector (and to fix the free parameter), we use the measurements of the amplitude of the scalar power spectrum $A_s$ and its running $n_s$. Those parameter shed light, in particular, on the cosmological behavior at the time $t_*$, when the pivot scale $k_*=0.05$ Mpc$^{-1}$ exited the horizon. At that time, the universe can be considered as flat \cite{Bonga:2016iuf} and we further assume that the slow-roll approximation is satisfied \cite{Martin:2013tda}, {\it i.e.}
\begin{align}
    \epsilon_*&:=-\frac{\dot{H}_*}{H_*^2}\simeq \frac{1}{16\pi}\left(\frac{V_{,\phi}(\phi_*)}{V(\phi_*)}\right)^2\ll 1\, ,
    \\
    \eta_*&:=\left.\dv{\ln{\epsilon}}{N}\right|_{t_*}\simeq 4 \,\epsilon_* -\frac{1}{4\pi}\frac{V_{,\phi\phi}(\phi_*)}{V(\phi_*)}\ll 1\, ,
\end{align}
where $N$ is the number of e-folds and the subscript ``$,\phi$" means a derivative with respect to the scalar field. Using the standard definitions of $A_s$ and $n_s$, together with the field equations at the time $t_*$, namely
\begin{align}
    &H_*\simeq \sqrt{\pi A_s \epsilon}\, ,\quad n_{s*}\simeq 1-2\epsilon_*-\eta_*\, ,\label{eq:def_As_ns}
\\
    &3 H_* \dot{\phi}_*+V_{,\phi}(\phi_*)\simeq 0\, ,\label{eq:slow_roll_raychaudhuri}
\\
    &H_*^2\simeq\frac{8\pi}{3}\left(\frac{\dot{\phi}_*^2}{2}+V(\phi_*)\right)\, ,\label{eq:slow_roll_friedmann}
\end{align}

one can solve for $\phi_*$, $\dot{\phi}_*$, and for the free parameter included in the definition of $V(\phi)$, either $m$ or $M$. It is then possible to choose $\phi(t=0)$ and $\dot{\phi}(t=0)$ at the time of the bounce so as to satisfy conditions \eqref{eq:def_As_ns}, \eqref{eq:slow_roll_raychaudhuri} and \eqref{eq:slow_roll_friedmann} at the time $t_*$.

Actually, because inflation is a strong attractor, some freedom is left in choosing the IC at the time of the bounce. To better understand the physical meaning of the IC, we discuss them using the energy density of the scalar field $\rho$ and the equation of state $w:=p/\rho$, instead of $\phi$ and $\dot{\phi}$. The density $\rho$ allows one to immediately read the energy scale of the considered phase, while $w$ quantifies how far away from a pure de Sitter space-time the universe is. In practice, whatever the chosen value at the bounce $w(t=0)\in [-1,-1/3]$, inflation does occur. Should the initial value of $w$ be higher than $-1/3$, the bounce would simply not happen. On the other hand, the energy density at the bounce $\rho(0)$ is constrained by the physical conditions at the time $t_*$, which in turn constrain the number of e-folds $N$ of inflation for a given model. As shown in details in \cite{Renevey:2020zdj}, when $w>-1/3$ the energy density of matter strongly dominates at the origin of our universe and the big bang singularity cannot be avoided. To be more precise, since $\dd w/\dd t\neq 0$ at $t=0$ and since the requirement $w<-1/3$ needs to be satisfied right before the bounce, the actual practical upper limit for $w(0)$ is slightly smaller than $-1/3$. Setting the IC for $\rho$ and $w$ is in fact equivalent to setting the IC for $\phi$ and $\abs*{\dot{\phi}}$. We therefore are also free to choose the sign of the derivative of the scalar field. The background IC are given by $\{w(0),\textrm{sgn}(\dot{\phi}(0))\}$, $\rho(0)$ being fixed by the conditions at $t_*$.

\begin{figure}
    \centering
    \includegraphics[width=0.9\linewidth]{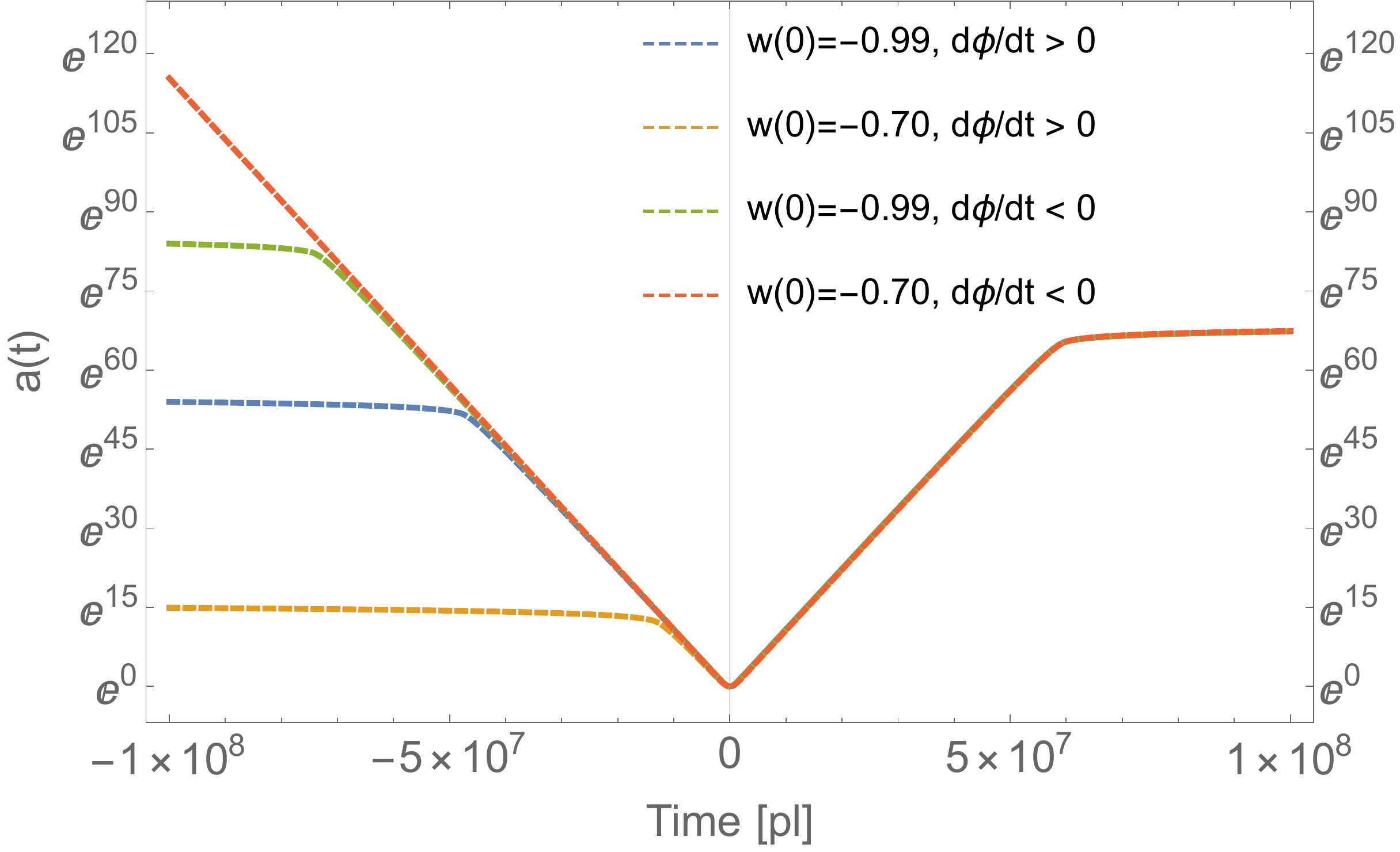}
    \caption{Evolution of the scale factor $a(t)$ before, during and after the bounce, for different initial conditions.}
    \label{fig:scale_factor_different_IC}
\end{figure}

Figure~\ref{fig:scale_factor_different_IC} illustrates the consequences of different choices of IC, by displaying the evolution of the scale factor during the pre-bounce phase and during the inflationary period (with the Starobinsky potential), for different choices of IC. First, one can indeed see that the choice of $w$ and $\textrm{sgn}(\dot{\phi})$ at the bounce has no influence on the inflationary period, as long as $\rho(0)$ is chosen accordingly. Remarkably, while the post-bounce behaviour is unaffected by the choice of IC, the pre-bounce universe, more precisely the duration of deflation, significantly is. For $\dot{\phi}(0)>0$ the deflationary period is shortened, while for $\dot{\phi}(0)<0$ it is lengthened. The choice of $w$ affects the asymmetry between deflation and inflation. Since deflation (just before the bounce) and inflation (just after it) are both required by the model, the scalar field lies high on its potential at the time of the bounce. The choice $\phi(0)>0$ can therefore be made without loss of generality. Furthermore, the derivative of the scalar field being non-vanishing at $t=0$, $\phi$ is not at its maximum value, but is instead slightly offset. This is the origin of the asymmetry between the length of deflation and the length of inflation. The quadratic potential leads to similar behaviours.\\

The critical aspect of this otherwise very appealing scenario -- built only on the standard model of cosmology together with a positive spatial curvature -- is obviously the deflation phase. It raises two important questions that are both physical and philosophical.

The first question is about the probability for this deflation phase to take place and last long enough. Deflation is unstable. If one thinks from the causal viewpoint of the pre-bounce phase, driving the universe to a de-Sitter contraction (the effective cosmological constant remaining positive) requires a large amount of fine-tuning. Some specific choices of potentials may overcome this difficulty \cite{Matsui:2019ygj} but, clearly, the deflationnary stage generally remains a repulsor. Is this a problem? Not necessarily. The {\it a priori} probability for most phenomena around us is extraordinary small. The viewpoint adopted here is the one of an archaeologist trying to figure out what has happened taking into account the laws we know and the facts we see. Not the one of a historian trying to determine the deep causes of a strange political event. There is nothing odd with discovering phenomena whose probability of occurrence was {\it a priori} incredibly small \cite{Barrau:2020nek}. The point is not to find an explanation for the path followed by the universe but to determine what this path might have been considering the knowledge we have.

The second question is related to the favored duration of deflation. Only a few e-folds are required for the model to work. This however raises an interesting point. If deflation was brief, it means that, when going backward in time prior to the bounce, the curvature term happened to dominate before the cosmological constant had a chance to overcome the dynamics. This implies that -- still thinking backward in time -- the universe was contracting again. The whole question of the singularity resolution therefore reappears. A new curvature bounce is obviously possible but still needs fine-tuning. The situation repeats itself as long as the deflation stages last only a few e-folds. On the other hand, if deflation was long enough (typically as long as inflation but the accurate value obviously depends on the unknown content of the contracting branch), the cosmological constant protects the pre-bounce phase from any re-contraction (once again in a time decreasing view). Only one bounce takes place. This might indicate that a single long period of deflation is somehow theoretically preferred. But this highly depends on arbitrary probability priors.\\

It is worth recalling here that, as we have shown in a previous article \cite{Barrau:2020nek}, the number of inflationary e-folds and the reheating temperature are fixed in this model. Consistency reasons impose $N\sim 70$ and $T_{RH}\sim T_{GUT}$. It is quite impressive that the number of e-folds precisely coincides with the minimum value required to account for observations and that the bounce energy scale (which is the same than the inflation scale) is substantially sub-Plankian, implying a well controlled classical behavior.

\section{Linear perturbations}

In order to compare the predictions of the model with the temperature anisotropies measured by Planck \cite{Aghanim:2018eyx}, one needs the theory of linear perturbations. In the following, we very briefly summarize the procedure for positively curved spaces \cite{Asgari:2014gra,Mukhanov:1990me}. On top of the homogeneous and isotropic background, small perturbations are added according to
\begin{align}
    g_{\mu\nu}\simeq \mathring{g}_{\mu\nu}+\delta g_{\mu\nu}
    \quad\textrm{;}\quad
    \phi\simeq \mathring{\phi}+\delta \phi\, ,
\end{align}
$\mathring{g}_{\mu\nu}$ and $\mathring{\phi}$ being the background quantities. As a symmetric rank two tensor, the metric perturbations can be decomposed into four scalar, four vector and two tensor degrees of freedom. Some of these degrees of freedom are gauge artifacts ans gauge-invariant quantities have to be constructed. In the case of scalar perturbations of the metric, this leads to the well-known Bardeen variables \cite{Bardeen:1980kt}, usually denoted $\Phi$ and $\Psi$. The tensor degrees of freedom are naturally gauge invariant. The matter fluctuations described by $\delta\phi$ have to be connected to the metric fluctuations through the Einstein field equations. Using the perturbed version of the equations, one can show that, for a perfect fluid having no anisotropic stress at linear order, the Bardeen variables are equal: $\Phi=\Psi$. The perturbed equations of motion also give the evolution equations for the two remaining scalar perturbations, linking $\Phi$ to $\delta\phi$. The quantization of the remaining scalar degree of freedom is performed in the canonical form using the Mukhanov-Sasaki (MS) variable \cite{Mukhanov:1985rz,Sasaki:1986hm}, which can be written as
\begin{align}
    v(t,\vb{x})=a\left(\delta\phi+\frac{\mathring{\phi}'}{\mathcal{H}}\Phi\right),
\end{align}
where a prime represents the derivative with respect to the conformal time $\eta$ (such that $d\eta = dt / a$) and $\mathcal{H}=a'/a$. The canonical scalar perturbation is directly related to the curvature perturbation $\mathcal{R}$, through
\begin{align}
    v=\frac{a\,\mathring{\phi}'}{\mathcal{H}}\mathcal{R}.
\end{align}

So far, we have worked in position space, where the fluctuations are functions of time and spatial coordinates $v(\eta,\vb{x})$. 
As opposed to the spatially flat case, where one describes functions of spatial coordinates using Fourier expansions, when the positive curvature is taken into account, the discrete basis of hyperspherical harmonics $Q_{nlm}(\vb{x})$ has to be used, since the manifold is closed. The detailed procedure is explained in \cite{Bonga:2016iuf}. In short, the hyperspherical harmonics can be separated into a radial part $f_{nl}(\chi)$ and a spherical harmonic $Y_{lm}(\theta,\varphi)$ part according to $Q_{nlm}(\vb{x})=f_{nl}(\chi)\cdot Y_{lm}(\theta,\varphi)$. Using this basis, any function defined on the 3-sphere, including the MS variable $v$, can be expanded as 
\begin{align}
    v(\eta,\vb{x})=\sums_{n=2}^{\infty}\sums_{l=0}^{n-1}\sums_{m=-l}^{l}v_{nlm}(\eta)Q_{nlm}(\vb{x})\, .
\end{align}
The perturbed field equations describing the behaviour of the MS variable can now be written in momentum space, leading to an equation of motion for each mode $v_{nlm}$, namely \cite{Bonga:2016iuf,Bonga:2016cje}
\begin{align}
    v_{nlm}''+A_n(\eta)v_{nlm}'+B_n(\eta)v_{nlm}=0\, ,\label{eq:eom_scalar_modes}
\end{align}
where
\begin{align}
    A_n(\eta)  =& \left(32 \pi   a^3 \dot{a} \dot{\phi} V_{\phi}(\phi) 
              +48 \pi  a^2 \dot{a}^2 \dot{\phi}^2 
         \right.\nonumber\\&\left.
         -8 \pi   a^2 \dot{\phi}^2 \left(8 \pi   a^2 \left(\dot{\phi}^2-2 V \right)
               +2K\right)\right)\times
               \nonumber\\
               &\times\left(2\dot{a} \left(2(n^2-4)\dot{a}^2 
              + 8 \pi  a^2 \dot{\phi}^2 \right)\right)^{-1}\, ,
              \label{eq:An}
\end{align}
and
\begin{align}
    &B_n(\eta) =  \frac{8\pi}{\dot{a}^2 \left(2 \left(n^2-4\right) 
                \dot{a}^2+8 \pi   a^2 \dot{\phi}^2\right)} 
                \nonumber\\ 
         &\times\Bigg[ \frac{\dot{a}^4 (n^2-4) \left( (n^2-1)K + a^2 V_{\phi\phi} \right)}{4\pi }
         \nonumber\\&
         + \left(4
n^2-7\right) a^3 \dot{a}^3 \dot{\phi} V_{\phi} 
\nonumber\\&
-\pi \frac{n^2-1}{n^2-4} a^4\dot{\phi}^4 
         \left[8 \pi  a^2 \left(\dot{\phi}^2+2 V\right)-6K\right] 
         \nonumber\\
      &+   (n^2-1) a^2 \dot{a}^2  \times
      \nonumber\\&
      \times\Bigg(  
            -6 \pi   \frac{n^2-5}{n^2-4} a^2\dot{\phi}^4+ 
             4 \pi a^2 \dot{\phi}^2 V+\frac{3K}{2} \dot{\phi}^2 + \frac{9}{2}
\dot{a}^2 \dot{\phi}^2 \Bigg) 
\nonumber\\
&+ a^3 \dot{a} \left[ a \dot{a} \dot{\phi}^2 V_{\phi\phi}+2 a \dot{a}
V_{\phi}^2+ 4 \pi a^2 \dot{\phi} V_{\phi} \left(\dot{\phi}^2+2 V\right)- K\dot{\phi} V_{\phi}  \right]
\Bigg]
\nonumber\\&
- H~\big(H+A_n(\eta)/a\big) - \frac{\ddot a}{a}\, ,
\label{eq:Bn}
\end{align}
Analogies with the flat case can easily be drawn. Usually, scalar functions are expanded using the basis $Q_{klm}(\vb{x})=j_l(k r)Y_{lm}(\theta,\varphi)$, where $j_l(k r)$ are the Bessel functions and $k\in\mathbb{R}$ spans a continuous spectrum. In fact, in the limit where $n$ is large and $\chi$ is small (that is when positively curved spatial sections are nearly flat), the radial part of the hyperspherical harmonics $f_{nl}(\chi)$ behaves as $j_l(k r)$. Moreover, when comparing the equations of motion for the scalar modes, one can identify
\begin{align}
    (n^2-1)K\rightarrow k^2.
\end{align}
Hence, $\sqrt{(n^2-1)K}$ can be understood as the curved space counterpart of the flat space wave number $k$. From now on, we define the ``flat case limit wave number" to be $k_n=\sqrt{(n^2-1)K}$.

\begin{figure}
    \centering
    \begin{minipage}[h]{1.0\linewidth}
    \includegraphics[width=0.9\linewidth]{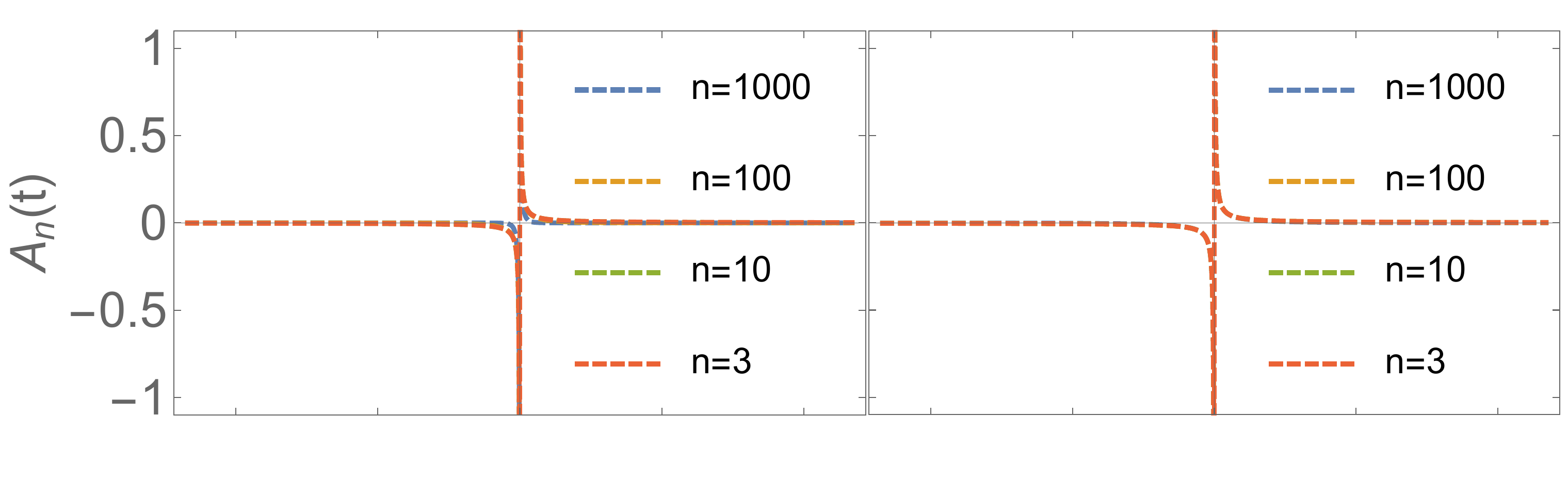}
    \end{minipage}
    
    \vspace{-0.3cm}
    
    \begin{minipage}[h]{1.0\linewidth}
    \includegraphics[width=0.9\linewidth]{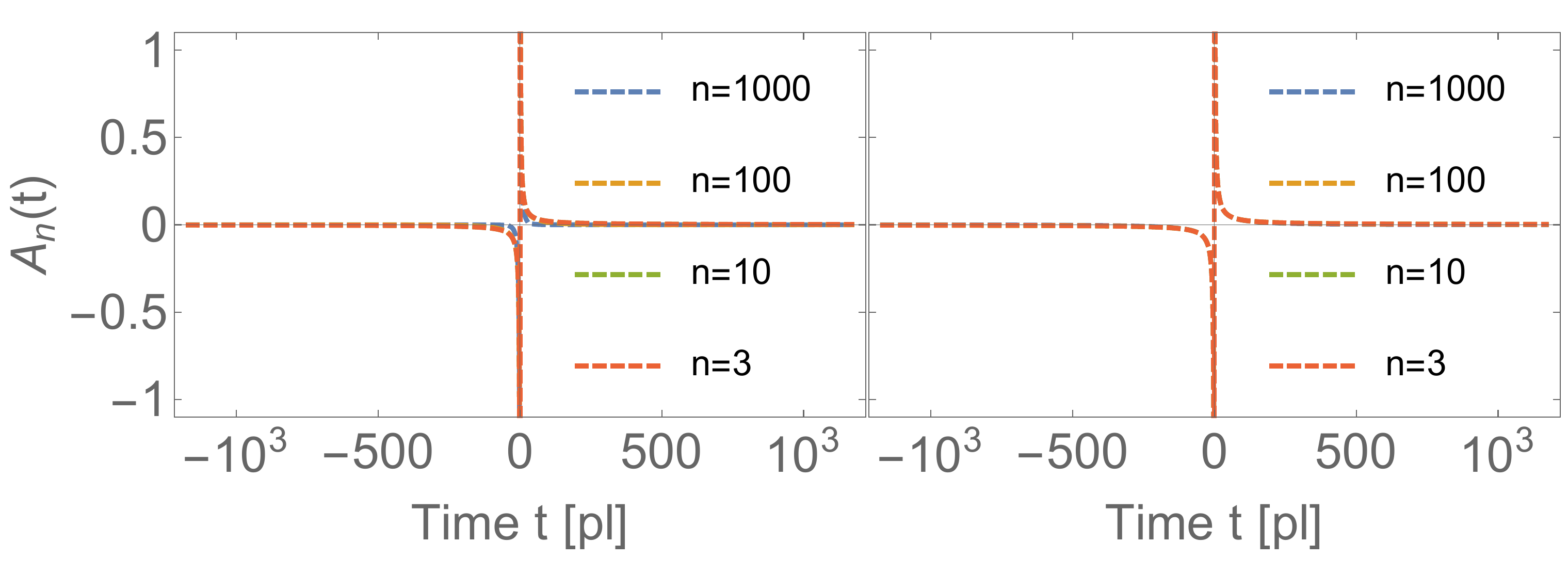}
    \end{minipage}
    \caption{Behaviour of the functions $A_n(t)$ around the bounce, $t=0$, for different wave numbers $n$ and initial conditions, in the case of a quadratic potential. \textit{top left:} evolution from the IC $\{-0.99,+1\}$, \textit{top right:} $\{-0.467,+1\}$, \textit{bottom left:} $\{-0.99,-1\}$, \textit{bottom right:} $\{-0.41,-1\}$.}
    \label{fig:An_quadratic}
\end{figure}

\begin{figure}
    \centering
    \begin{minipage}[h]{1.0\linewidth}
    \includegraphics[width=0.9\linewidth]{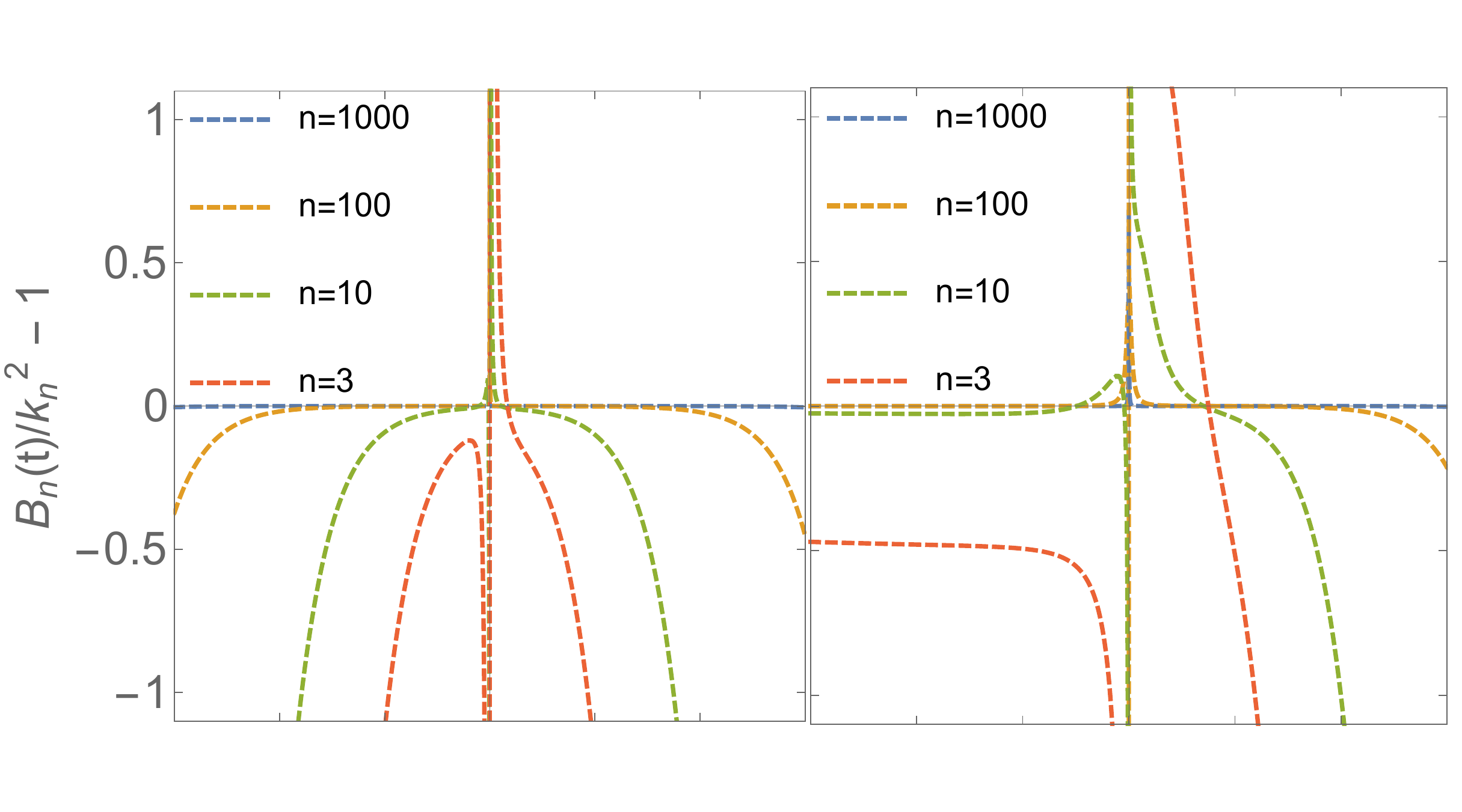}
        \end{minipage}
    
    \vspace{-0.6cm}
    
    \begin{minipage}[h]{1.0\linewidth}
    \includegraphics[width=0.9\linewidth]{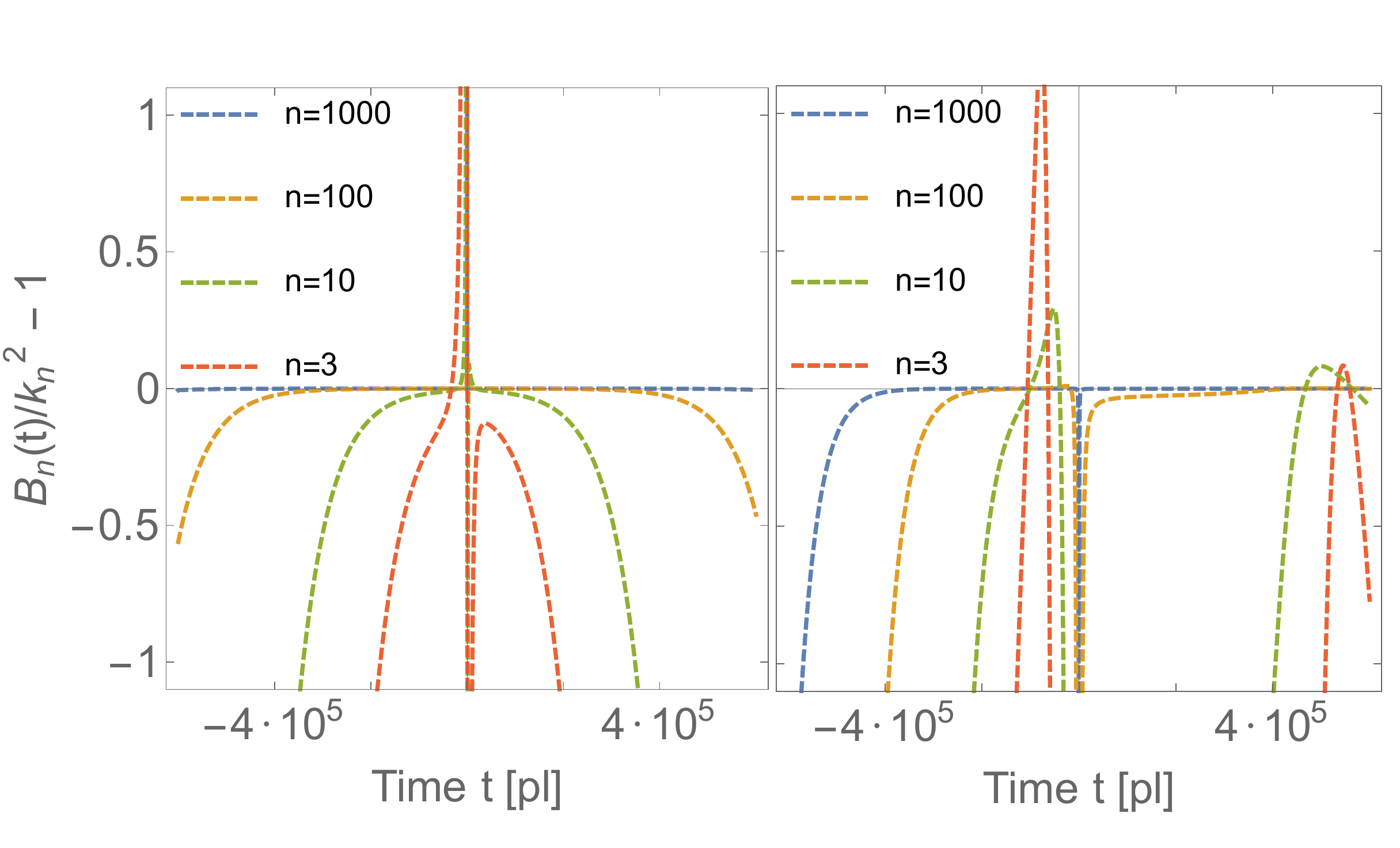}
    \end{minipage}
    \caption{Behaviour of the functions $B_n(t)$ around the bounce $t=0$, for different wave numbers $n$ and initial conditions, in the case of a quadratic potential. \textit{top left:} $\{-0.99,+1\}$, \textit{top right:} $\{-0.467,+1\}$, \textit{bottom left:} $\{-0.99,-1\}$, \textit{bottom right:} $\{-0.41,-1\}$.}
    \label{fig:Bn_quadratic}
\end{figure}

\begin{figure}
    \centering
    \begin{minipage}[h]{1.0\linewidth}
    \includegraphics[width=0.9\linewidth]{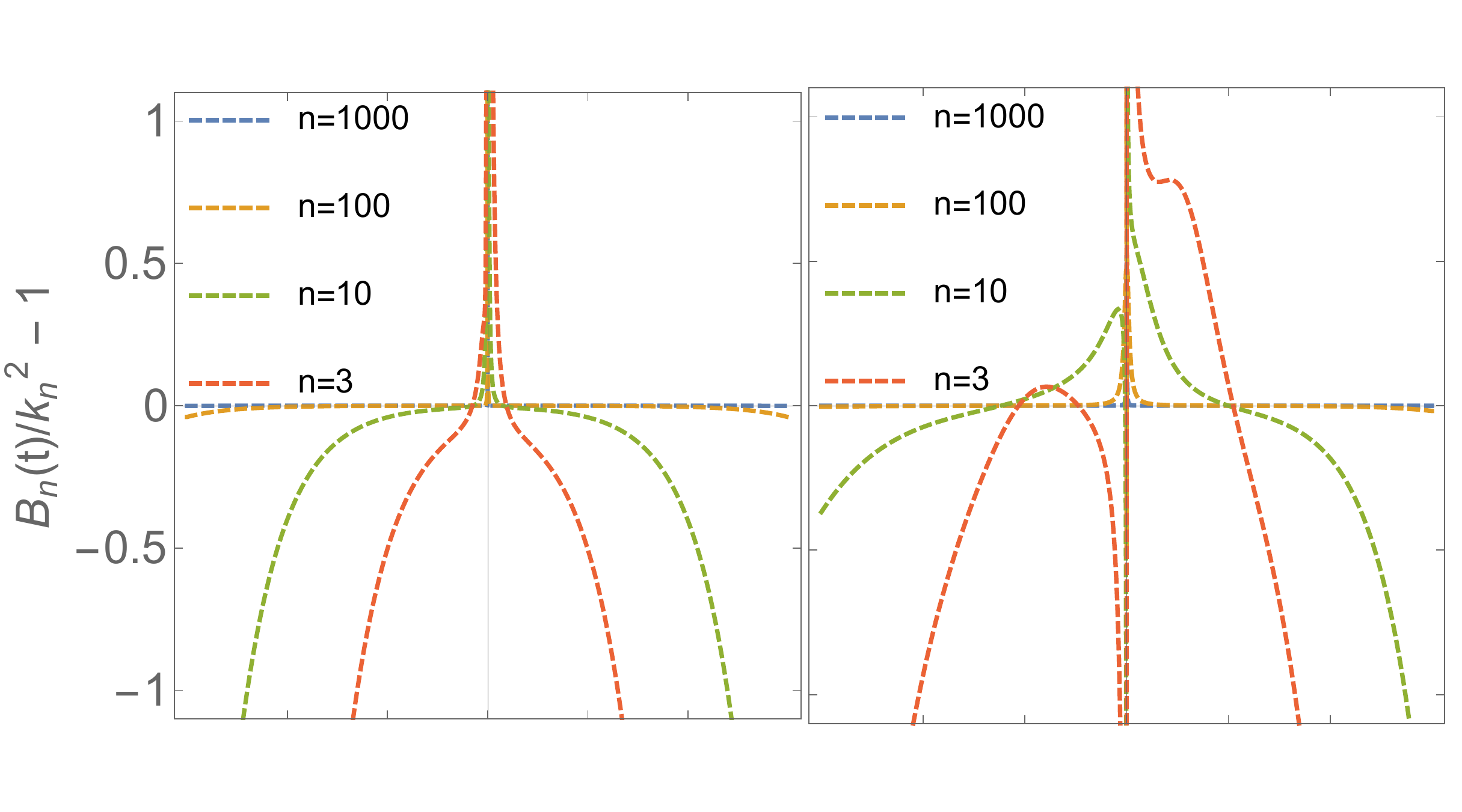}
    \end{minipage}
    
    \vspace{-0.6cm}
    
    \begin{minipage}[h]{1.0\linewidth}
    \includegraphics[width=0.9\linewidth]{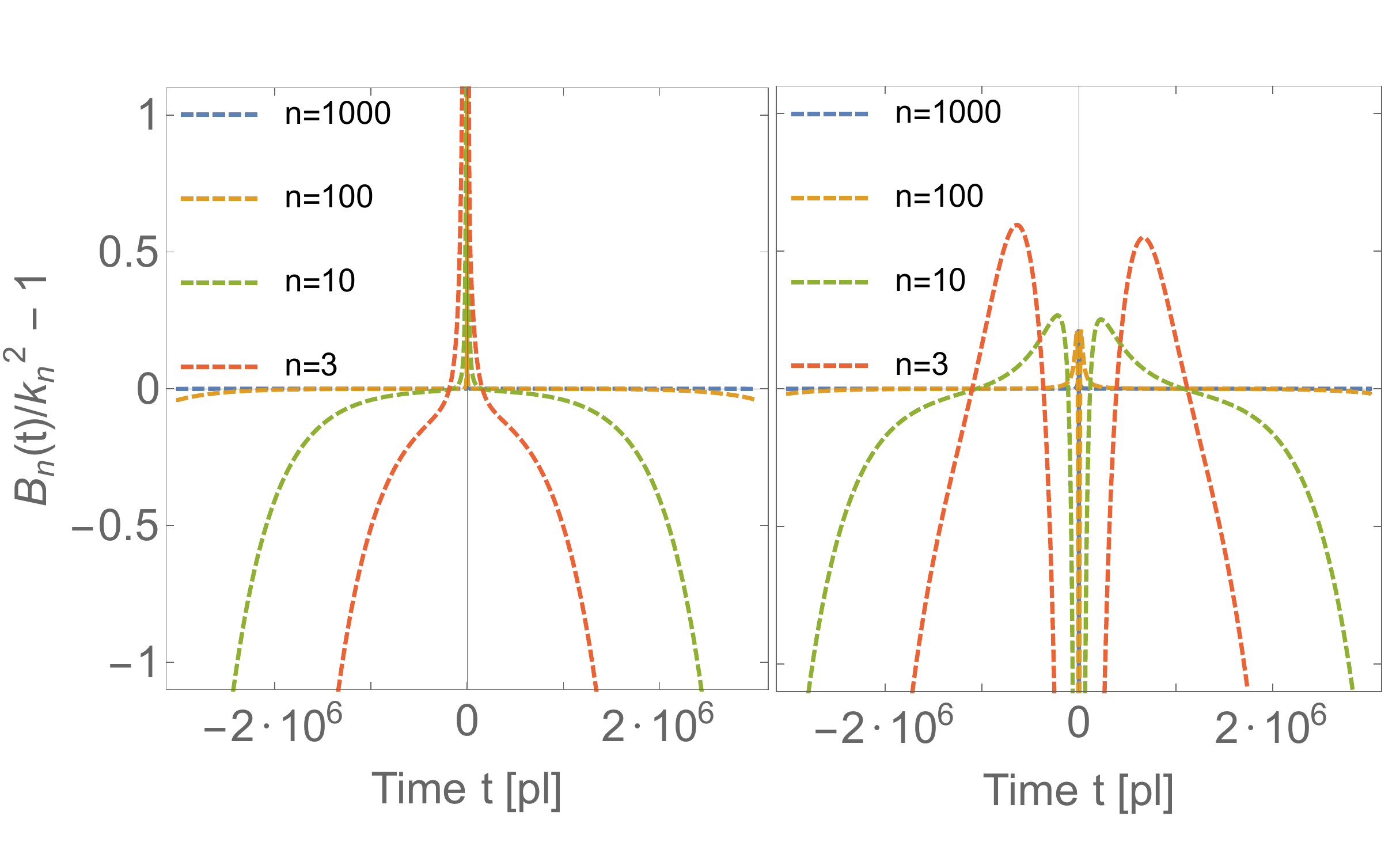}
    \end{minipage}
    \caption{Behaviour of the functions $B_n(t)$ around the bounce $t=0$, for different wave numbers $n$ and initial conditions, in the case of a Starobinsky potential. \textit{top left:} evolution from the IC $\{-0.99,+1\}$, \textit{top right:} $\{-0.5,+1\}$, \textit{bottom left:} $\{-0.99,-1\}$, \textit{bottom right:} $\{-0.5,-1\}$.}
    \label{fig:Bn_starobinsky}
\end{figure}

The scalar perturbations are to be understood as due to quantum fluctuations taking place close to the bounce. To implement this, we follow the standard quantization techniques of cosmological perturbations, setting the initial quantum state for $v_n(\eta)$ in the so-called Bunch-Davies vacuum. It is unique and equivalent to the minimum energy state if and only if the quantum fluctuations behave as in Minkowski space-time, {\it i.e.} $A_n\rightarrow 0$ and $B_n\rightarrow k_n^2$, at the initial time $t_i$ \cite{Schander:2015eja}. Those requirements are not trivial in the considered scenario, as we shall see later on. When they are fulfilled, the IC for the perturbations can be set as 
\begin{align}
    v(t_i)=\frac{1}{k_n}\quad\textrm{\&}\quad v'(t_i)=-i\,\frac{k_n}{\sqrt{2}}.
\end{align}
To check the validity of the Bunch-Davies vacuum approximation, the functions $A_n(t)$ for a massive inflaton are drawn in Fig.~\ref{fig:An_quadratic}, for different values of $\{w(0),\textrm{sgn}(\dot{\phi}(0))\}$ and $n$. The ``deviations", $B_n(t)/k_n^2 -1$ are shown in Fig.~\ref{fig:Bn_quadratic}. The case of a Starobinsky potential is considered in Fig.~\ref{fig:Bn_starobinsky}. We do not show the functions $A_n(t)$ in the latter case as they are similar to those of the quadratic potential. The values $w(0)=-0.467$ and  $w(0)=-0.5$, respectively for the quadratic and starobinsky potentials, with $\dot{\phi}(0)>0$, are chosen to minimize the number of e-folds of deflation. It is as low as only $1$ to $2$ e-folds in both cases (we recall that the bounce cannot occur without deflation). For a negative scalar derivative $\dot{\phi}(0)<0$, the values $w(0)=-0.41$ and $w(0)=-0.5$ are chosen, leading to a very high number of deflationnary e-folds: $185$ and $760$, respectively, for the quadratic and the Starobinsky potentials. Clearly, in all cases, the functions $A_n(t)$ and $B_n(t)$ are non-trivial and even divergent at the bounce. This is, partly, why when the fluctuations are initiated before the bounce, they might leave a significant imprint on the scalar perturbations and consequently on the scalar power spectrum. Furthermore, the functions $A_n(t)$ become very small close the bounce, therefore satisfying the first requirement for the Bunch-Davies vacuum. It should also be noticed that they do not significantly vary with respect to the wave number $n$ and to the shape of the inflaton potential. The second requirement, $B_n(t_i)\approx k_n^2$, is harder to fulfill for small values of $n$, as it can be seen in Figs~\ref{fig:Bn_quadratic} and \ref{fig:Bn_starobinsky} -- even more so for higher values of $w(0)$. In particular, in the case of the quadratic potential with IC $\{-0.467,+1\}$, the Minkowski vacuum can hardly be met with a precision better than $50$\%. Although this is {\it not} problematic for the model itself, it makes the clear analysis of the power spectra less obvious. For the vast majority of IC, it however remains possible to define a time $t_i$ before or after the bounce, such that $B_n$ is close enough to the desired value $k_n^2$. For larger wave numbers, {\it e.g.} $n\geq 10$, the Bunch-Davies vacuum is always very well defined for an appropriate choice of IC.

It should be stressed that, in this model, the Bunch-Davies vacuum can only be chosen for a limited amount of time, either before or after the bounce. We assume that it makes sense to select the minimum energy state when the instantaneous Minkowski vacuum is met. Although reasonable and usual, this is not a straightforward assumption.


Beyond the specific model considered in this study, the question of the origin of cosmological perturbations in bouncing models is a difficult one. For example, in the case of loop quantum gravity, some authors advocate the idea that IC should be implemented at the bounce \cite{Agullo2}, whereas others insist on putting them in the contracting branch \cite{Bolliet:2015bka}. This raises questions both on the practical side and on the conceptual side. Practically, the issue is the one of the most convenient way to proceed. Conceptually, the question is the one of causality. If time flows in the same direction before and after the bounce\footnote{This hypothesis is not obvious. If, as suggested by Penrose \cite{Penrose:1980ge}, irreversibility is driven by processes that takes place while approaching (nearly) singular points, it could be that time -- in this particular sense (as there are actually many different times at play \cite{Rovelli:2021elq}) -- flows in two opposite directions from the bounce.}, and if the word ``initial" is meant literally, the IC should be defined before the bounce. Of course, the detailed underlying process, leading to the inflaton domination (anti-reheating) is still to be described. It is our opinion that, even at the heuristic level, it is meaningful to initiate the perturbations before the bounce, at the time when the Bunch-Davis vacuum is approached as closely as possible.

\section{Primordial power spectrum and temperature anisotropies}

In previous sections, we have discussed the background behaviour leading to a curvature bounce instead of the big-bang singularity and explained how the cosmological evolution depends upon the IC. We have also introduced the theory of cosmological perturbations in a closed universe and showed that quantum fluctuations can be initiated either before or after the bounce, using in all cases the Bunch-Davies vacuum. The primordial scalar power spectrum and the subsequent effects on the temperature anisotropies measured in the CMB can now be calculated for different hypotheses that will be described later. 

To compute the primordial power spectrum, we first evolve the perturbations $v_n(t)$ from the initial time $t_i$ to the reheating using the equation of motion \eqref{eq:eom_scalar_modes}. At the reheating, the curvature term is completely negligible when compared to the other fluids and one can use the standard definition of the power spectrum of the scalar perturbations, namely
\begin{align}
    \mathcal{P}_S(n)=\left.\frac{k_n^3}{2\pi^2}\abs{\frac{v_n}{z}}^2\right|_{t=t_{rh}},
\end{align}
$t_{rh}$ being the reheating time. 

To calculate the spectrum of temperature anisotropies, we use the Boltzmann code \textbf{CAMB}, which is well suited to take into account a positive curvature and a discrete primordial power spectrum \cite{Lewis:1999bs}.

\subsection{Post-bounce origin of structures}

We start with the assumption that the anisotropies observed in the CMB originates from quantum fluctuations generated after the bounce. This situation is nearly analogous to the one considered in the work of~\cite{Bonga:2016iuf} (where no bounce occurs), with the exception of the numerical value of curvature parameter ($\Omega_K=-0.005$ in their case). The specific time at which the IC for the background and for the perturbations are set is also different. One of the main results of~\cite{Bonga:2016iuf}  was the prediction of a decrease of the primordial power spectrum amplitude at low values of $n$ and, consequently, a decrease of the temperature anisotropies at low multipolar numbers $\ell$. In~\cite{Bonga:2016iuf}, the background IC are chosen at the time $t_*$, when the modes with $k_*=0.005$ Mpc$^{-1}$ exited the horizon. The scale factor and the scalar field are then evolved both in the future, $t>t_*$, and in the past, $t<t_*$. Although fully legitimate, this approach might miss the ``standard" behaviour as the slow-roll solution is not an attractor when going backward in time (which is equivalent to going forward in time in a contracting universe). This leads to instabilities (the same than those responsible for the low likelihood of a long lasting deflation in the contracting branch) and to the big bang singularity. The fluctuations are set in the Bunch-Davies vacuum before the time $t_*$, when all the modes are sub-Hubble. However, at that time, the requirements of the Bunch-Davies vacuum are in fact not fully satisfied for low wave numbers $n$ and it might very well be that this is the reason for the calculated damping of the primordial power spectrum at low multipolar numbers.

\begin{figure}
    \centering
    \begin{minipage}[h]{1.0\linewidth}
    \includegraphics[width=0.9\linewidth]{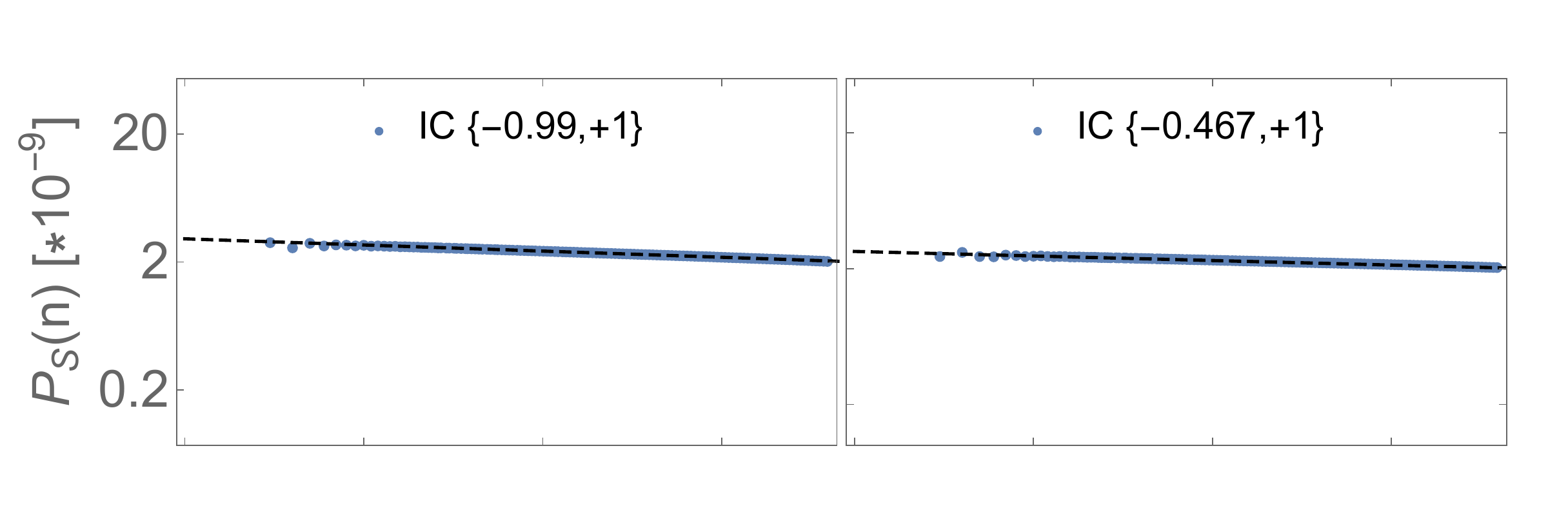}
    \end{minipage}
    
    \vspace{-0.6cm}
    
    \begin{minipage}[h]{1.0\linewidth}
    \includegraphics[width=0.9\linewidth]{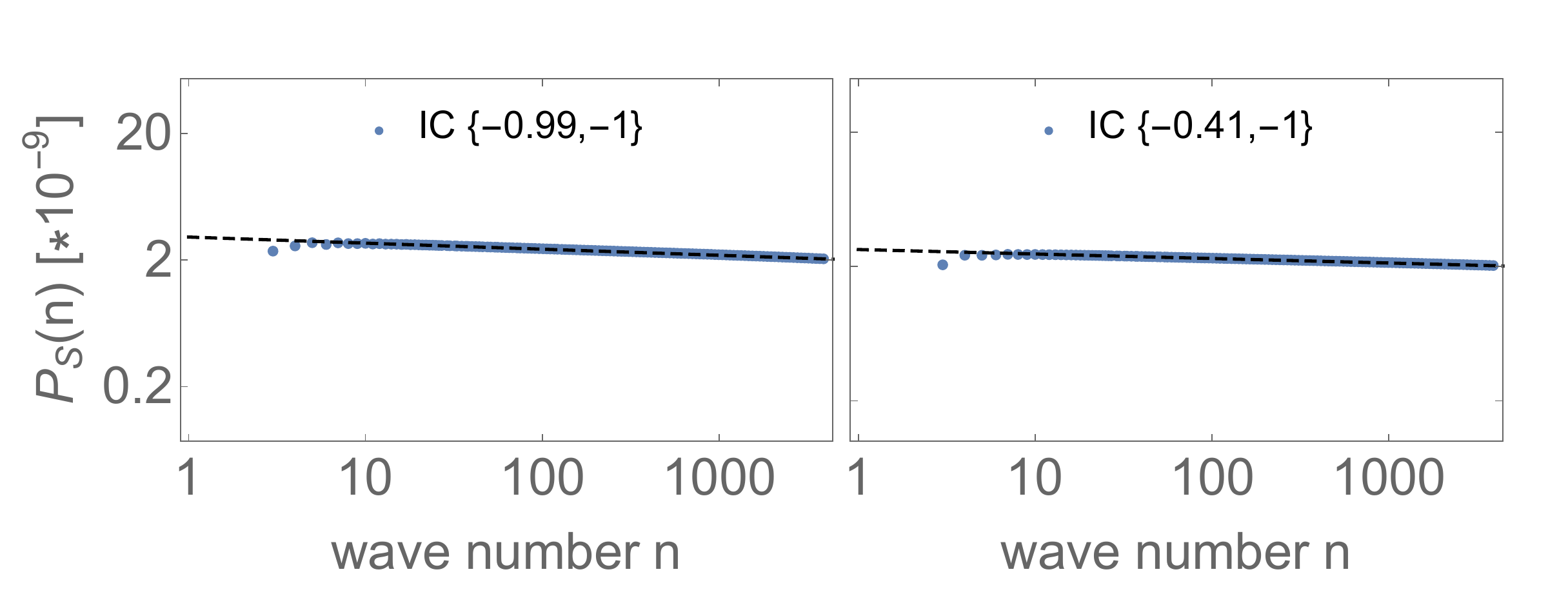}
    \end{minipage}
    \caption{Primordial scalar power spectrum for different set of background initial conditions, for a quadratic potential, and with fluctuations initiated after the bounce (blue dots). The standard power law is represented as a dashed black line. The IC for the background are: \textit{top left} $\{-0.99,+1\}$, \textit{top right} $\{-0.467,+1\}$, \textit{bottom left} $\{-0.99,-1\}$, \textit{bottom right} $\{-0.41,-1\}$.}
    \label{fig:PPS_quadratic}
\end{figure}
\begin{figure}
    \centering
    \begin{minipage}[h]{1.0\linewidth}
    \includegraphics[width=0.9\linewidth]{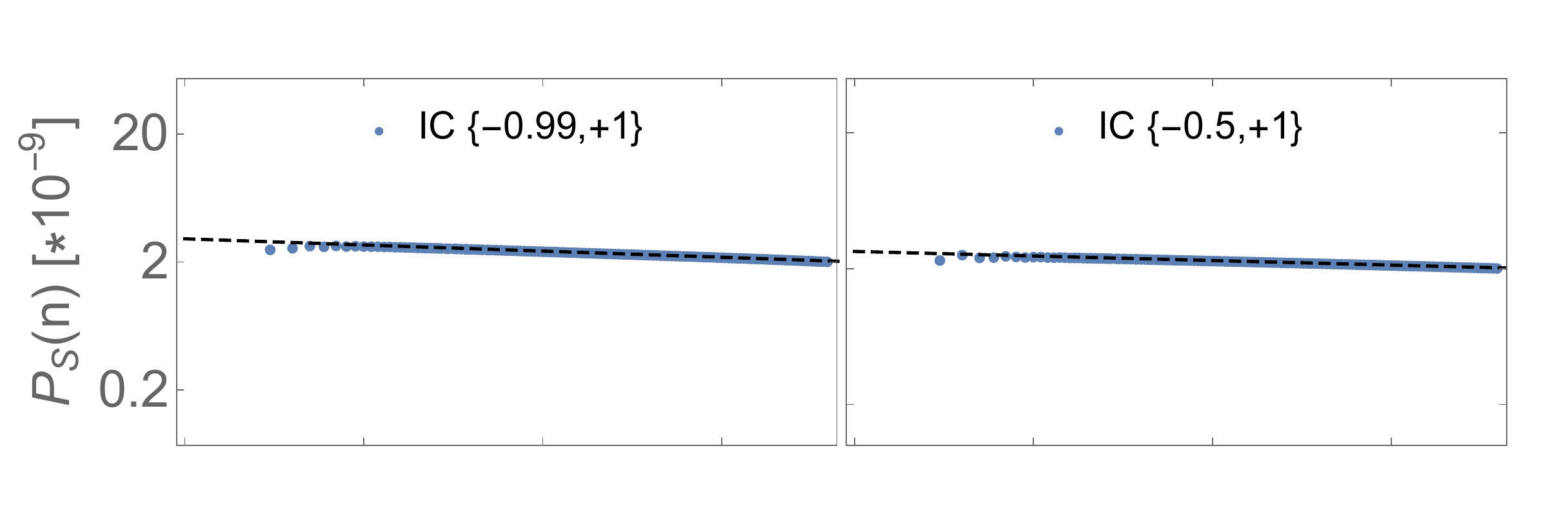}
    \end{minipage}
    
    \vspace{-0.6cm}
    
    \begin{minipage}[h]{1.0\linewidth}
    \includegraphics[width=0.9\linewidth]{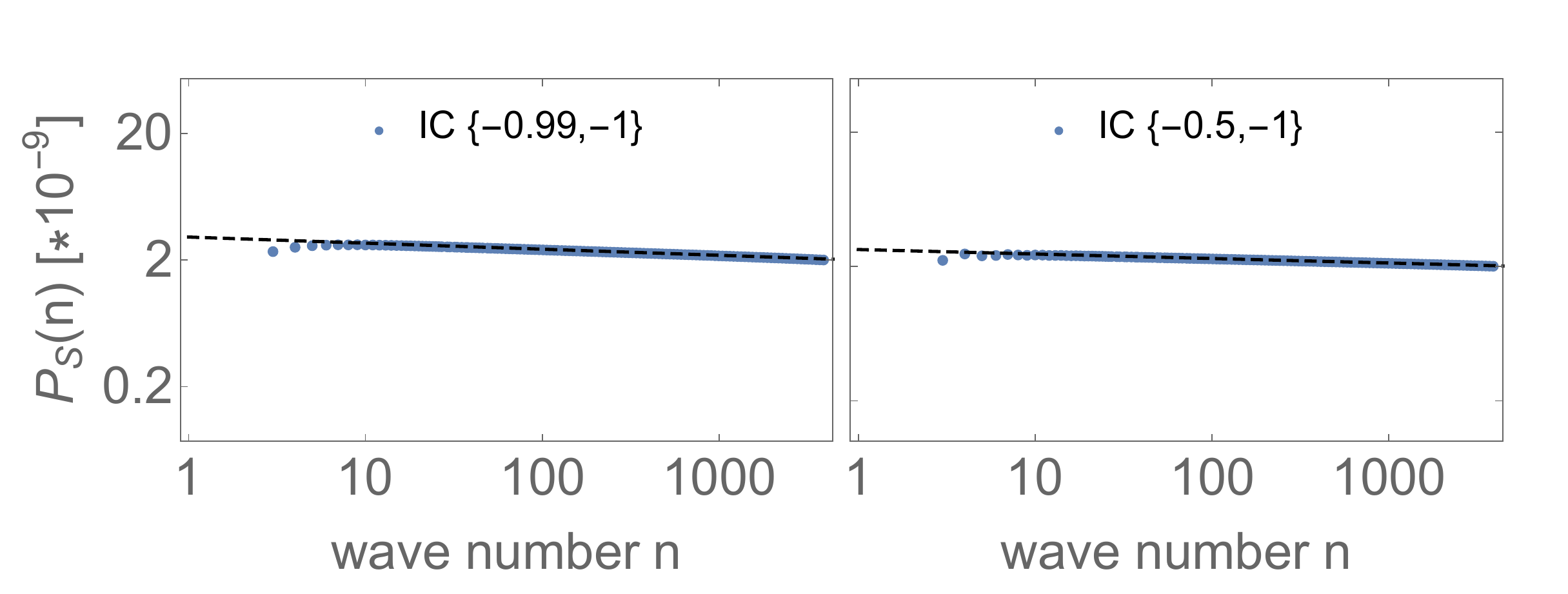}
    \end{minipage}
    \caption{Primordial scalar power spectrum for different set of background initial conditions, for a Starobinsky potential, and with fluctuations initiated after the bounce (blue dots). The standard power law is represented as a dashed black line. The IC for the background are: \textit{top left} $\{-0.99,+1\}$, \textit{top right} $\{-0.5,+1\}$, \textit{bottom left} $\{-0.99,-1\}$, \textit{bottom right} $\{-0.5,-1\}$.}
    \label{fig:PPS_starobinsky}
\end{figure}

\begin{figure}
    \centering
    \includegraphics[width=0.9\linewidth]{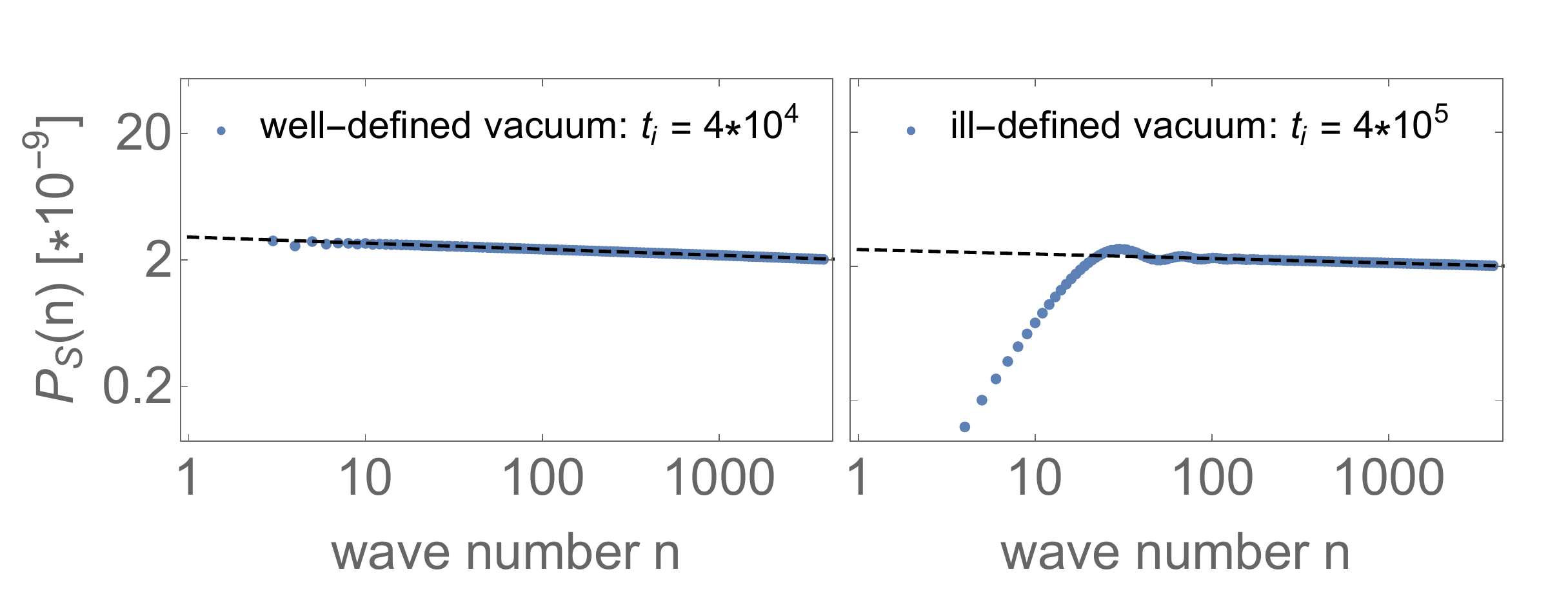}
    \caption{Comparison between the primordial power spectra of scalar modes from a well-defined (left) and ill-defined (right) Bunch-Davies vacuum. The calculation is performed for the quadratic potential with IC $\{-0.99,+1\}$.}
    \label{fig:PPS_quadratic_wrong}
\end{figure}

To support this last argument, we have calculated the primordial scalar power spectrum for both potentials and for different values of the background IC. In each case, we select the best post-bounce time $t_i>0$, where the requirements for a Bunch-Davies vacuum are satisfied, to initiate the quantum fluctuations. The results for the quadratic and Starobinsky potentials are exposed in Fig.~\ref{fig:PPS_quadratic} and Fig.~\ref{fig:PPS_starobinsky}, respectively. It can be seen that, if one chooses the most appropriate time to set the IC for the perturbations, the primordial power spectra for the scalar modes in a positively curved space are almost identical to those of a flat universe, described by as a power law $\mathcal{P}_S(k)=A_s(k/k_*)^{n_s-1}$. Even for the IC $\{-0.99,-1\}$, where the Minkowski vacuum conditions are not exactly met, Fig.~\ref{fig:Bn_quadratic} shows that one closely recovers the flat space power spectrum. To further support that the lack of power of the primordial power spectrum at low $n$ found in \cite{Bonga:2016iuf} might be due to an ill-defined Bunch-Davies vacuum, one can calculate the scalar power spectrum with a sub-optimal initial time $t_i$. The comparison between the power spectra initiated from a well-defined and an ill-defined Bunch-Davies vacuum for the quadratic potential with background IC $\{-0.99,+1\}$ is show on Fig.~\ref{fig:PPS_quadratic_wrong}. A significant deficit of power for the ill-defined vacuum, similar to the one of \cite{Bonga:2016iuf}, is indeed observed. We do not show the temperature anisotropies for the primordial power spectra shown in Figs.~\ref{fig:PPS_quadratic} and \ref{fig:PPS_starobinsky}, since they are the same as in the flat space case.

\subsection{Pre-bounce origin of quantum fluctuations}

In most cases, the Minkowski vacuum requirements are met {\it before} the curvature bounce and it is reasonable to investigate the situation where the quantum fluctuations are initiated in their minimum energy state at that time. If this assumption is correct, one has to evolve perturbations through the bounce -- the later leaving a characteristic imprint. To compute the primordial power spectrum, we proceed as before, {\it i.e.} choosing the best initial time $t_i<0$ to set the IC for the perturbations -- that is corresponding to a good quality vacuum. They are then propagated up to the end of inflation and the power spectrum is calculated. This procedure is repeated for both potentials and for different background IC. The divergence of the functions $A_n$ and $B_n$ is an additional numerical difficulty.  Fortunately, the divergence being well localized and well behaved, one can always find a time interval around $t=0$, where both functions can be accurately approximated by $A_n(t)\approx a_n/t$ and $B_n(t)-k^2\approx b_n/t$, where $a_n$ and $b_n$ can be well determined. With these approximations, the equation for the perturbations \eqref{eq:eom_scalar_modes} can be solved analytically using Bessel functions. Outside this specific time interval around the bounce, the perturbations are evolved with numerical techniques. 

\begin{figure}
    \centering
    \begin{minipage}[h]{1.0\linewidth}
    \includegraphics[width=0.9\linewidth]{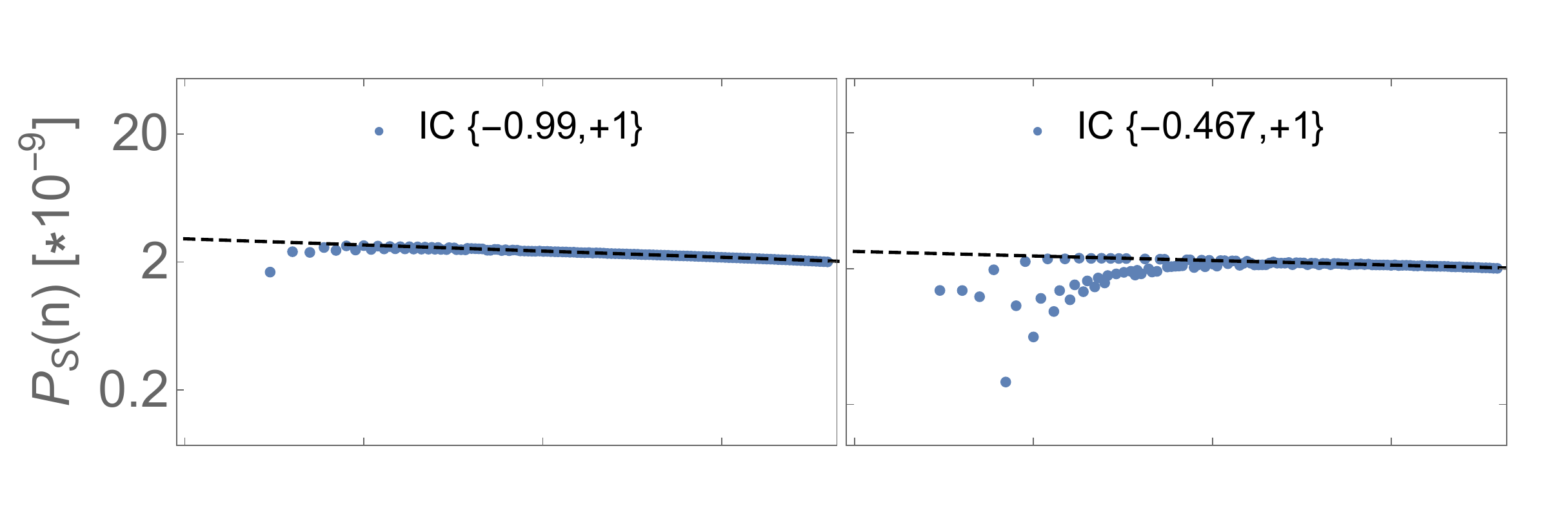}
    \end{minipage}
    
    \vspace{-0.6cm}
    
    \begin{minipage}[h]{1.0\linewidth}
    \includegraphics[width=0.9\linewidth]{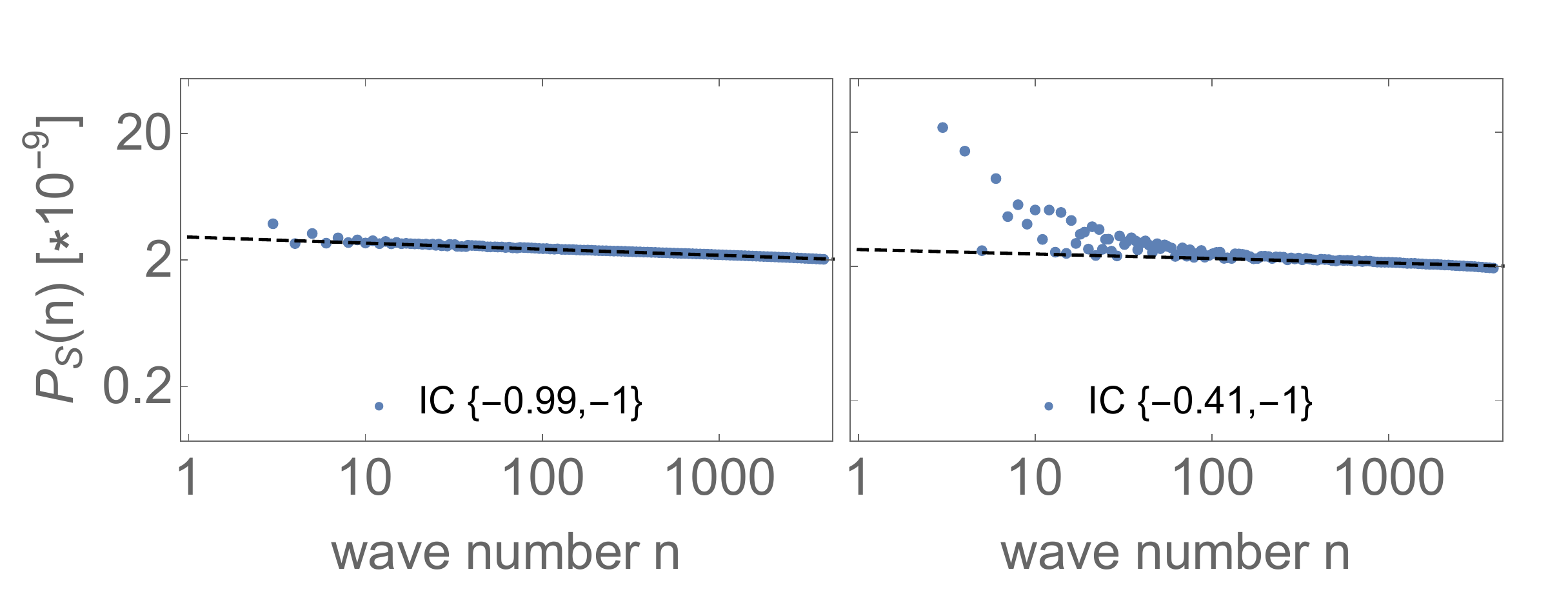}
    \end{minipage}
    \caption{Primordial scalar power spectrum for different sets of background initial conditions, in the case of the quadratic potential and with fluctuations initiated before the bounce (blue dots). The dashed black line represents the standard power law. The IC for the background are: \textit{top left} $\{-0.99,+1\}$, \textit{top right} $\{-0.467,+1\}$, \textit{bottom left} $\{-0.99,-1\}$, \textit{bottom right} $\{-0.41,-1\}$.}
    \label{fig:PPS_quadratic_pre}
\end{figure}
\begin{figure}
    \centering
    \begin{minipage}[h]{1.0\linewidth}
    \includegraphics[width=0.9\linewidth]{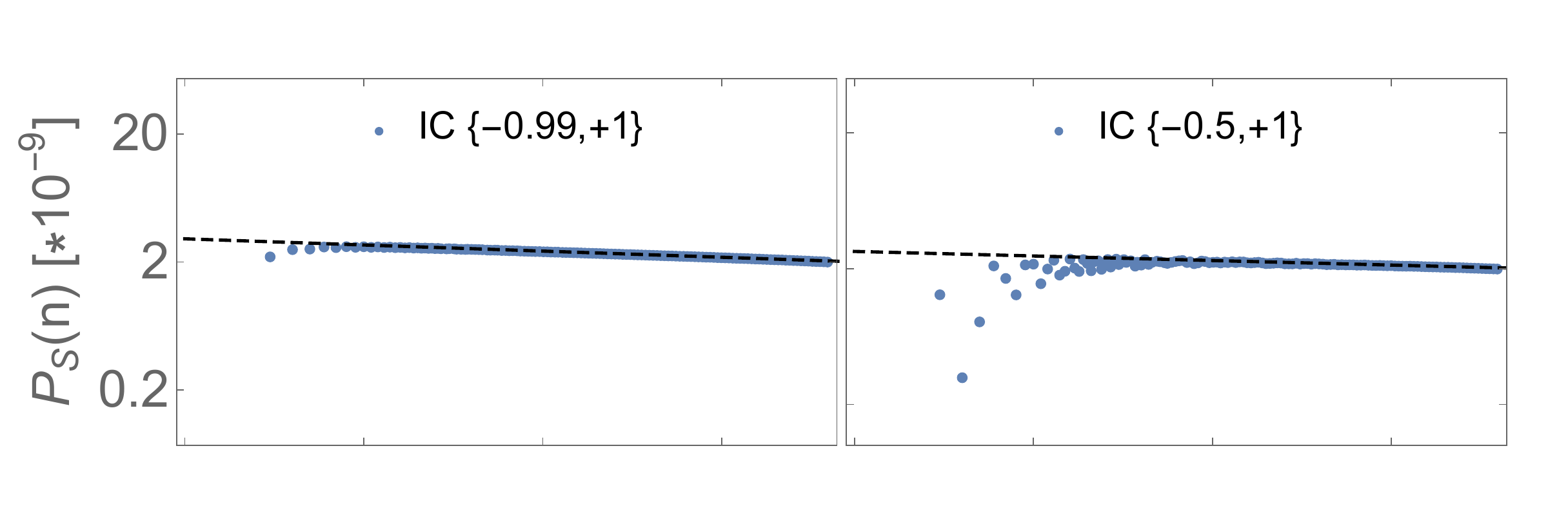}
    \end{minipage}
    
    \vspace{-0.6cm}
    
    \begin{minipage}[h]{1.0\linewidth}
    \includegraphics[width=0.9\linewidth]{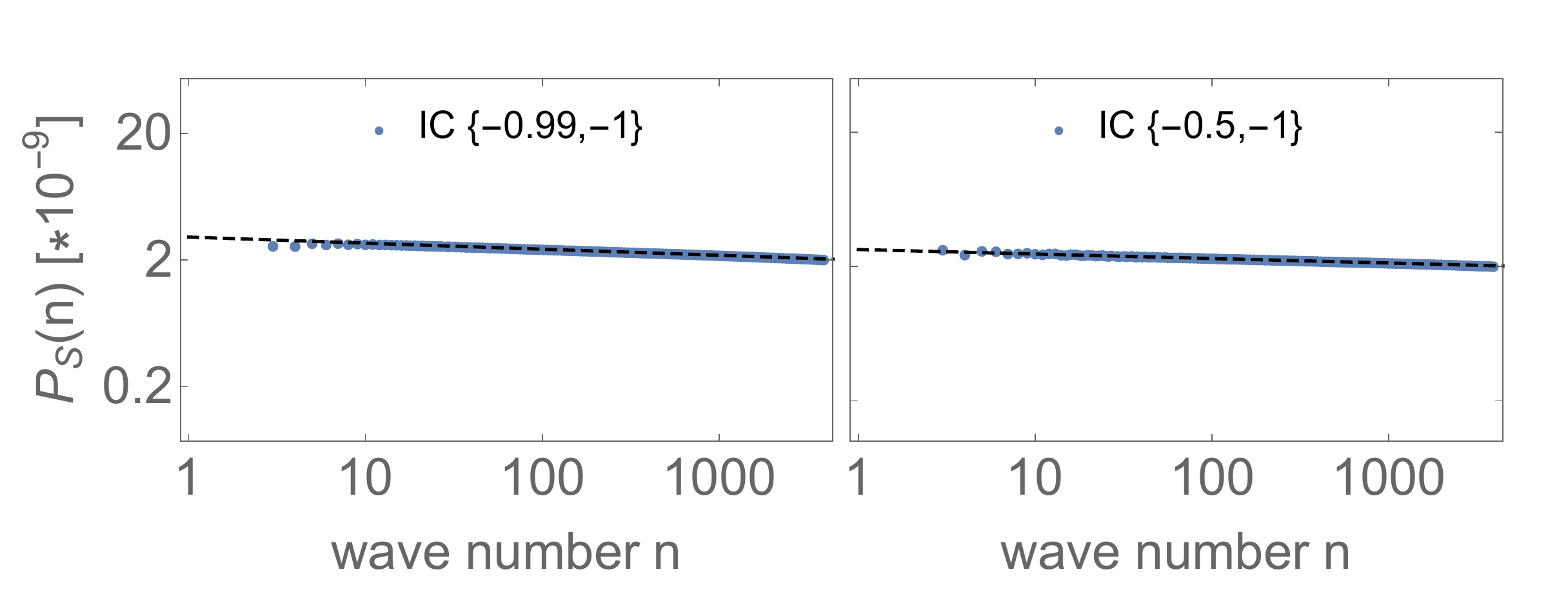}
    \end{minipage}
    \caption{Primordial scalar power spectrum for different sets of background initial conditions, in the case of the quadratic potential and with fluctuations initiated before the bounce (blue dots). The dashed black line represents the standard power law. The IC for the background are: \textit{top left} $\{-0.99,+1\}$, \textit{top right} $\{-0.5,+1\}$, \textit{bottom left} $\{-0.99,-1\}$, \textit{bottom right} $\{-0.5,-1\}$.}
    \label{fig:PPS_starobinsky_pre}
\end{figure}

The Bunch-Davies vacuum cannot be properly defined in the case of the quadratic potential with background IC $\{-0.467,+1\}$, hence we expect a decline in the primordial scalar power spectrum at low wave numbers. For the other power spectra, we  expect the shape to be only modified by the bounce. The primordial power spectrum for scalar fluctuations initiated before the curvature bounce, in the case of the quadratic and Starobinsky potentials, are shown in Figs.~\ref{fig:PPS_quadratic_pre} and \ref{fig:PPS_starobinsky_pre} respectively. For both potentials, when the duration of the deflation is similar to the duration of inflation, represented by $w(0)=-0.99$, it can be seen that the primordial power spectrum is not significantly affected. As expected, for a massive inflaton generating a very short period of deflation, corresponding to IC $\{-0.467,+1\}$, we indeed observe a decline in the power spectrum amplitude at low wave numbers. However, we also observe this behaviour in the case of the Starobinsky potential, for which the vacuum is well-defined. It can then be safely conjectured that the short period of deflation is the origin of the lack of power at low $n$. When the period of deflation is longer than the period of inflation, the result significantly differ, depending on the potential. For the massive scalar field, the primordial power spectrum at low $n$ is amplified when compared to the flat space case, while for the Starobinsky scalar field, it is not drastically affected.

\begin{figure}
    \centering
    \includegraphics[width=0.99\linewidth]{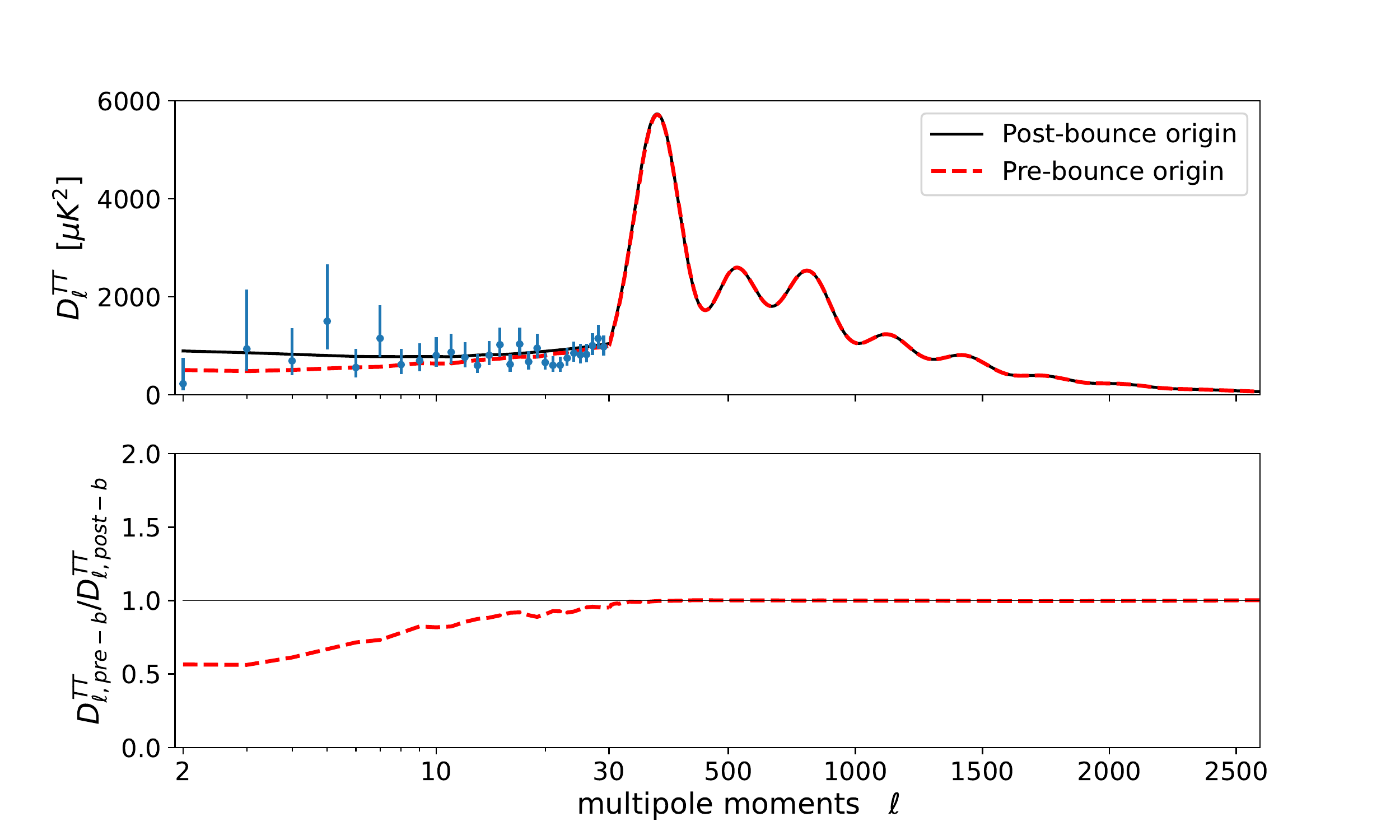}
        \caption{Temperature angular power spectra $D_\ell^{\textrm{TT}}$ for a pre-bounce and for a post-bounce origin of structures, with background IC $\{-0.467,+1\}$, and a quadratic potential. \textit{Top:} $D_{L,\textrm{post-b}}^{\textrm{TT}}$ in black, $D_{\ell,\textrm{pre-b}}^{\textrm{TT}}$ in dashed red, Planck data in blue. \textit{Bottom:} Ratio of the spectra in dashed red.}
    \label{fig:DL_quadratic_positive}
\end{figure}
\begin{figure}
    \centering
    \includegraphics[width=0.99\linewidth]{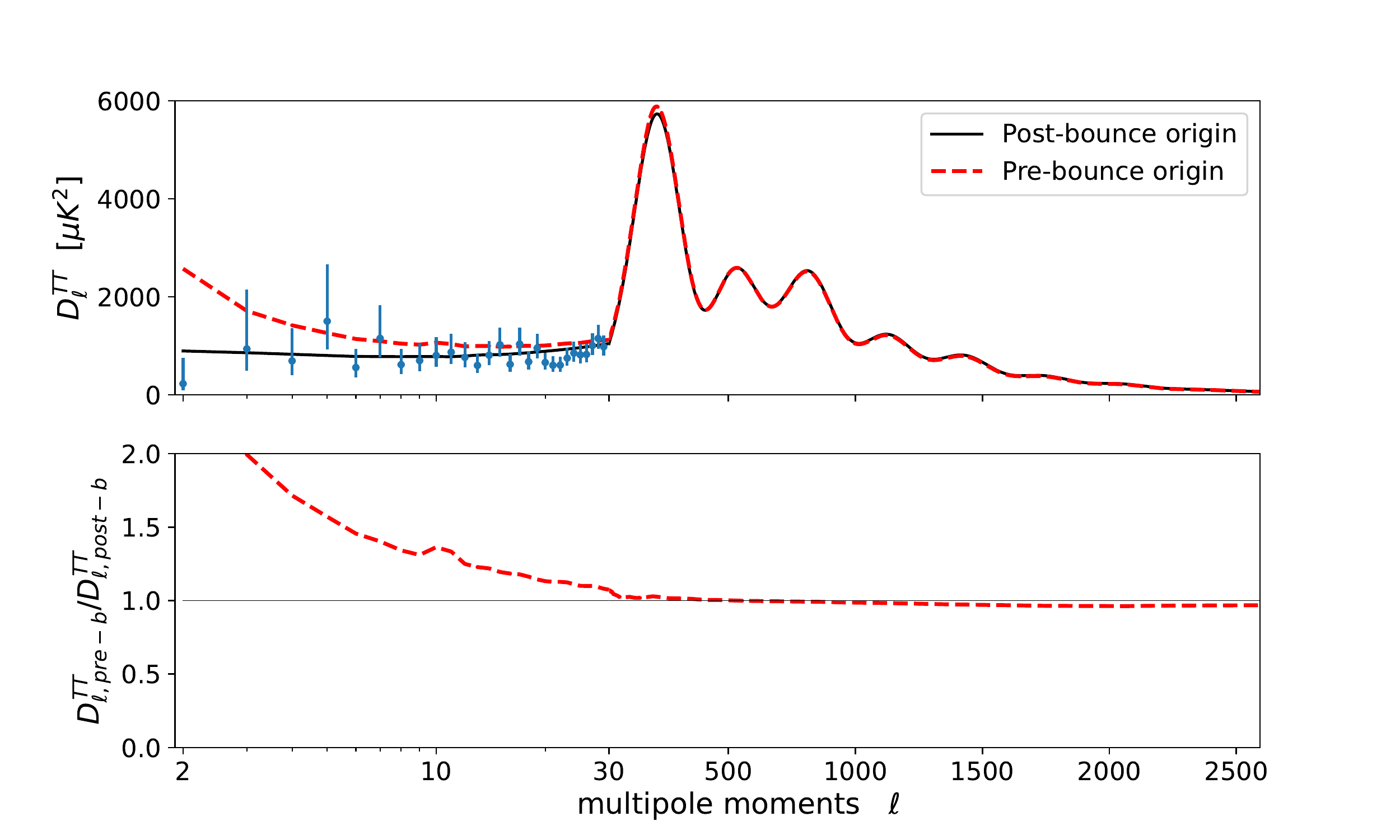}
    \caption{Temperature angular power spectra $D_\ell^{\textrm{TT}}$ for a pre-bounce and for a post-bounce origin of structures, with background IC $\{-0.41,-1\}$, and quadratic potential. \textit{Top:} $D_{\ell,\textrm{post-b}}^{\textrm{TT}}$ in black, $D_{\ell,\textrm{pre-b}}^{\textrm{TT}}$ in dashed red, Planck data in blue. \textit{Bottom:} Ratio of the spectra in dashed red.}
    \label{fig:DL_quadratic_negative}
\end{figure}
\begin{figure}
    \centering
    \includegraphics[width=0.99\linewidth]{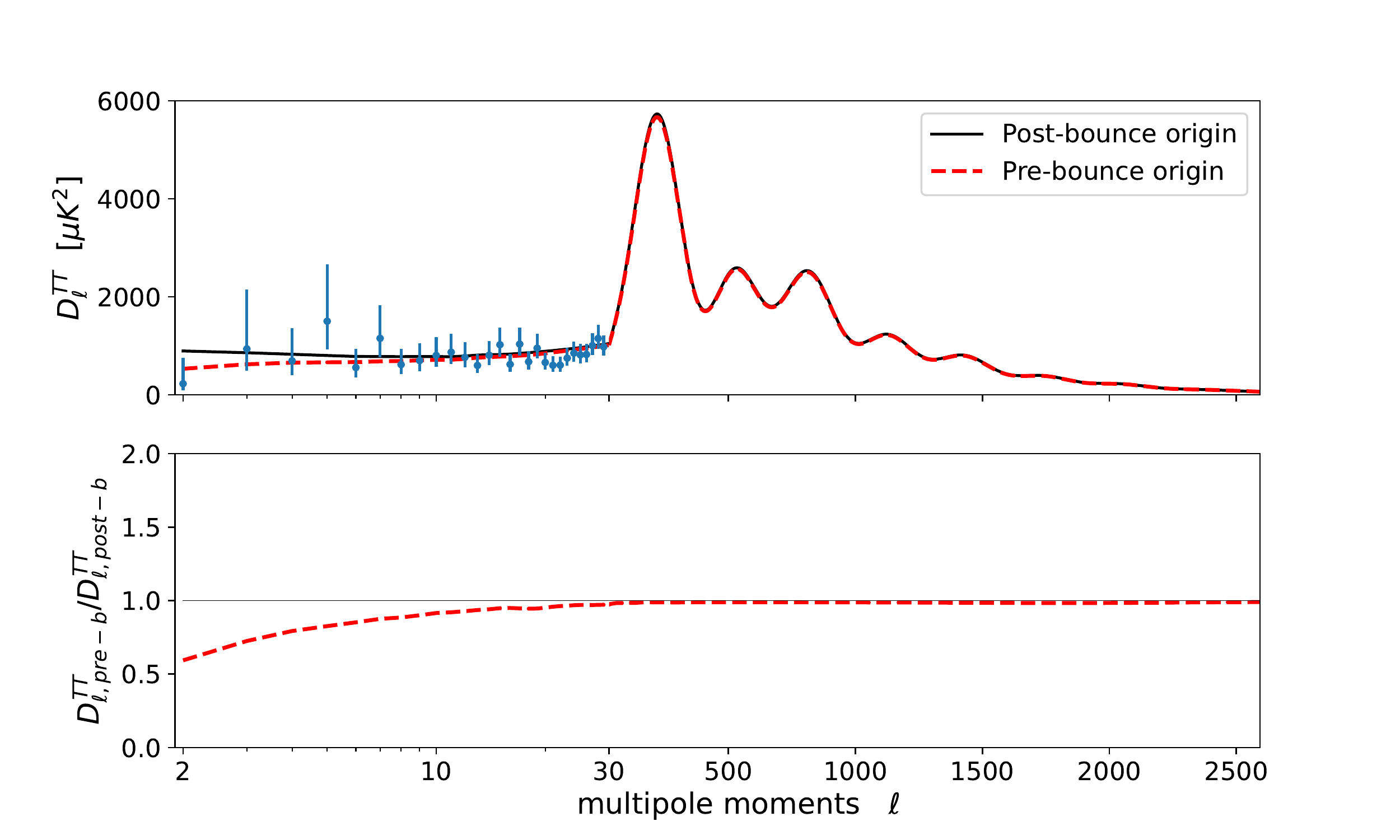}
    \caption{Temperature angular power spectra $D_\ell^{\textrm{TT}}$ for a pre-bounce and for a post-bounce origin of structures, with background IC $\{-0.5,+1\}$, and Starobinsky potential. \textit{Top:} $D_{\ell,\textrm{post-b}}^{\textrm{TT}}$ in black, $D_{\ell,\textrm{pre-b}}^{\textrm{TT}}$ in dashed red,  Planck data in blue. \textit{Bottom:} Ratio of the spectra in dashed red.}
    \label{fig:DL_starobinsky_positive}
\end{figure}

Three of the eight primordial scalar power spectra deviate from the standard power law $\mathcal{P}_S(k)=A_s(k/k_*)^{n_s-1}$, hence we focus on those for the calculation of the temperature  power spectra $C_L^{\textrm{TT}}$. The deviation mainly appears at low wave number $n$ and the $C_\ell^{\textrm{TT}}$ are therefore affected at low values of $\ell$, where the cosmic variance  makes it difficult to have statistically significant measurements. Nonetheless, strong deviations could lead to inconsistencies. Although not statistically significant, it might also be argue that explaining the slight lack of power observed in the CMB at large scales \cite{Bunn:2008zd} would be welcome. To calculate the $C_\ell^{\textrm{TT}}$, or more precisely the $D_\ell^{\textrm{TT}}:=\ell(\ell+1)/(2\pi)C_\ell^{\textrm{TT}}$, for the three different non-trivial cases, namely quadratic potential with IC $\{-0.467,+1\}$ and $\{-0.41,-1\}$, and the Starobinsky potential with IC $\{-0.5,+1\}$, we proceed as follow. We use the Boltzmann code \textbf{CAMB}, modifying the cosmological parameters for the spatial curvature and for the Hubble constant according to \cite{DiValentino:2019qzk}. With these appropriate values and the standard power law for the primordial power spectrum, one recovers almost exactly the prediction made with the usual parameters $\Omega_K=0$ and $H_0=67.3$ km/s/Mpc, as found in \cite{Aghanim:2018eyx} (this is why, in \cite{DiValentino:2019qzk}, it can be argued that a positively curved universe is compatible with CMB measurements). We then compute two different power spectra.  The first one, named $D_{\ell,\textrm{post-b}}^{\textrm{TT}}$, is calculated using the standard power law for the primordial spectrum, which will serve as a comparison. As the name suggests, it is equivalent to the result calculated for the primordial power spectra derived with post-bounce initial conditions. The second temperature power spectrum $D_{\ell,\textrm{pre-b}}^{\textrm{TT}}$ is computed with the modified primordial power spectra from the pre-bounce initial perturbations. 

The temperature power spectra for the quadratic potential with a short period of deflation, a long period of deflation, and for the Starobinsky potential with a short period of deflation are shown in Figs.~\ref{fig:DL_quadratic_positive}, \ref{fig:DL_quadratic_negative} and \ref{fig:DL_starobinsky_positive}, respectively. While the important increase of power at low values of $\ell$, as can be observed in Fig.~\ref{fig:DL_quadratic_negative} for a long period of deflation, is disfavored by Planck measurements, the slight decrease at low $\ell$ for the short deflation fits well the data. It should be underlined that we present here the extreme cases only (either an identical duration for deflation and inflation, or a very long/brief period of deflation). One can of course choose background IC in between these extremes and obtain a weaker deviation. Overall, Planck data do slightly favor a shorter period of deflation than inflation. When compared to the post-bounce initial conditions, the pre-bounce initial conditions with a short period of deflation decrease the $\chi^2$ per degree of freedom by, respectively, 2.25 and 1.40 for a quadratic potential and a Starobinsky potential (performing the analysis with points before the first acoustic peak only). Those numbers are to be considered as indications of the trend and not treated as statistically significant results. 

Overall, the curvature bounce hypothesis agrees with current data. Depending on the IC, it can fit slightly better or slightly worst the CMB observations, when compared to the standard model. For the vast majority of IC, it is hardly distinguishable from the usual scenario, which makes it reliable and consistent but hard to falsify. Some non-CMB ideas for experimental probes were however suggested in \cite{Barrau:2020nek}.

\section{Conclusion}

In the first part of this article, we have discussed the background behaviour in the pre-bounce universe, at the bounce and during the inflationary period.  In particular, we have shown that widely different pre-bounce dynamics can lead to the same inflationary stage, due to the strong attractor status of the slow-roll inflation trajectory for the background equations of motion. To be more precise, the duration of the period of deflation can be chosen freely, without affecting the physics in the expanding universe at the background level. This raises interesting theoretical questions. From the viewpoint of stability for a single bounce, a brief deflation period is favored. But from the viewpoint of the avoidance of multiple bounces, a long deflation is favoured. It should, once again, be emphasized that this work does not focus on the ``naturalness" of the considered cosmic evolution. The other way round, it tries to determine the correct solution -- whatever its {\it a priori} probability -- taking into account what we know about the Universe.

After having introduced the theory of linear perturbations in curved space and shown that the quantum fluctuations can be initiated in the Bunch-Davies vacuum either before or after the bounce, we have calculated the primordial power spectrum of scalar perturbations for both the quadratic and Starobinsky potentials, for different values of the duration of the deflation stage. We have reached the conclusion that when the initial conditions for the fluctuations are set after the bounce, the standard power law spectrum, $\mathcal{P}_S(k)=A_s(k/k_*)^{n_s-1}$, derived from the flat space case, is recovered nearly exactly. This result is in tension with \cite{Bonga:2016iuf} where a power decrease at low wave numbers was pointed out. We believe  that the difference is (maybe partially) due to the fact that the requirements for the Bunch-Davies vacuum are not satisfied at the time when fluctuations are initiated in \cite{Bonga:2016iuf} . 

Later on, we have explicitly computed the primordial power spectrum for perturbations that are initially set before the bounce. We have found that, in the case of the quadratic potential, the results significantly differ from the standard power law, when the deflation period is either very long (typically around $185$ e-folds) or very short, (around $1$ or $2$ e-folds). The primordial power spectrum decreases at low wave numbers $n$ for a very short deflation, which is partly explained by an ill-defined Bunch-Davies vacuum. On the other hand, the power is increased at low $n$ for a very long deflation. In the case of the Starobinsky potential, the power spectrum is attenuated at low $n$ for a short deflation, but it is equivalent to the standard power spectrum for a long deflation, as opposed to the massive case. These modified primordial power spectra only affect the temperature  power spectrum $C_\ell^{TT}$ at low values of $\ell$, where it is difficult to reach a definitive conclusions due to the cosmic variance. Nonetheless, Planck data slightly favor a short duration of deflation, due to the power decrease at low wave numbers.

It is rather remarkable that, while the pre-bounce dynamics has no effects on the background behaviour of the expanding universe, the duration of the period of deflation can affect the temperature anisotropies observed in the CMB. To reach this conclusion, one needs to assume that fluctuations originate from the Bunch-Davies vacuum before the bounce. While it is a plausible hypothesis, this is obvioulsy not the only possible choice. In addition, in this work, we have extensively used the values of the parameters given in \cite{DiValentino:2019qzk} but other possibilities should be considered in future studies. 

Finally, it is worth recalling that this entire scenario is also appealing because the density of the Universe never approaches the Planck density. The bounce takes place for $\rho\ll \rho_{Pl}$ and the use of standard quantum field theory techniques is therefore much safer than in quantum gravity scenarios.

\bibliography{refs}

\end{document}